\documentclass[a4paper,11pt,twoside]{article}

\usepackage{geometry} 
\usepackage{amsmath}
\usepackage{amssymb}
\usepackage{color}
\usepackage{graphicx}
\usepackage{epstopdf}  % to use eps with Typeset -> Pdftex
\usepackage[]{units} % contains \nicefrac
\usepackage{bbold} % to obtain the identity matrix as \mathbb{1} (and R, C etc)
\usepackage{amsthm}
\usepackage{hyperref}
\usepackage{cite}

\definecolor{darkblue}{cmyk}{1,0.5,0,0.2}
\hypersetup{colorlinks, bookmarksnumbered, citecolor=darkblue, linkcolor=darkblue, pdfstartview=FitH, urlcolor=darkblue, linktocpage}

\makeatletter
\renewcommand{\section}{\@startsection{section}{1}{0em}%
        {-3.5ex \@plus -1ex \@minus -.2ex}% 
        {2.3ex \@plus.2ex}%
        {\normalfont\large\bfseries}}
\renewcommand{\subsection}{\@startsection{subsection}{2}{0em}%
        {-3.25ex\@plus -1ex \@minus -.2ex}%
        {1.5ex \@plus .2ex}%
        {\normalfont\bfseries}}
\renewcommand{\subsubsection}%
        {\@startsection{subsubsection}{3}{0em}%
        {-3.25ex\@plus -1ex \@minus -.2ex}%
        {1.5ex \@plus .2ex}%
        {\normalfont\itshape}}
\makeatother

\newcommand{\eV}{\,\mathrm{eV}}
\newcommand{\fracwithdelims}[4]{\left#1 \frac{#3}{#4} \right#2}
\newcommand{\ord}[1]{\mathcal{O}\left( #1 \right)}

\newcommand{\dm}[1]{{\Delta m^2_{#1}}}
\newcommand{\Fig}[1]{Fig.~\ref{fig:#1}}

\newcommand{\Tab}[1]{Tab.~\ref{tab:#1}}

\newcommand{\Sec}[1]{Sec.~\ref{sec:#1}}

\newcommand{\App}[1]{App.~\ref{sec:#1}}

\newcommand{\Eq}[1]{Eq.~(\ref{eq:#1})}
\newcommand{\eq}[1]{eq.~(\ref{eq:#1})}
\newcommand{\Eqs}[1]{Eqs.~(\ref{eq:#1})}
\newcommand{\eqs}[1]{eqs.~(\ref{eq:#1})}
\newcommand{\omi}[1]{}

\DeclareMathOperator{\diag}{Diag}
\DeclareMathOperator{\FT}{FT}

\newlength{\myem}
\newcommand{\sep}[1]{#1}
\newcounter{mysubequation}[equation]
\renewcommand{\themysubequation}{\alph{mysubequation}}
\newcommand{\mytag}{\stepcounter{mysubequation}%
\tag{\theequation\protect\sep{\themysubequation}}}
\newcommand{\globallabel}[1]{\refstepcounter{equation}\label{#1}}

\theoremstyle{definition}

\newtheorem*{lemma*}{Lemma}
\newtheorem*{conjecture*}{Conjecture}

\newtheorem*{proposition*}{Proposition}

\theoremstyle{definition}

\theoremstyle{remark}

\newcommand{\mtot}{m_\text{tot}}

\usepackage{pifont}% http://ctan.org/pkg/pifont
\newcommand{\cmark}{\ding{51}}%
\newcommand{\xmark}{\ding{55}}%

\newcommand{\SISSA}{SISSA/ISAS and INFN, I--34136 Trieste, Italy}
\newcommand{\ICTP}{ICTP, Strada Costiera 11, I--34151 Trieste, Italy}

\newcommand{\preprintdate}{April 2016}
\newcommand{\preprintnumber}{%
SISSA--25/2016/FISI}
 
\newcommand{\titletext}{Stable lepton mass matrices} 
\newcommand{\authortext}{\large Valerie Domcke$^{\, a}$ and Andrea Romanino$^{\, b,c}$
\medskip\\\em\normalsize 
$\mbox{}^a$ AstroParticule et Cosmologie (APC)/Paris Centre for Cosmological Physics, Universit\'e Paris Diderot, Paris, France
\\[0.1\baselineskip] 
$\mbox{}^b$ \SISSA 
\\[0.1\baselineskip] 
$\mbox{}^c$ \ICTP}
\newcommand{\abstracttext}{We study natural lepton mass matrices, obtained assuming the stability of physical flavour observables with respect to the variations of individual matrix elements. We identify all four possible stable neutrino textures from algebraic conditions on their entries. Two of them turn out to be uniquely associated to specific neutrino mass patterns. We then concentrate on the semi-degenerate pattern, corresponding to an overall neutrino mass scale within the reach of future experiments. In this context we show that i) the neutrino and charged lepton mixings and mass matrices are largely constrained by the requirement of stability, ii) naturalness considerations give a mild preference for the Majorana phase most relevant for neutrinoless double-$\beta$ decay, $\alpha \sim \pi/2$, and iii) SU(5) unification allows to extend the implications of stability to the down quark sector. The above considerations would benefit from an experimental determination of the PMNS ratio $|U_{32}/U_{31}|$, i.e.\ of the Dirac phase~$\delta$. }

\title{
\normalsize
\begin{tabular}[t]{l}%\hepnumber\\
\preprintdate\end{tabular}
\hspace*{\fill}
\begin{tabular}[t]{l}\preprintnumber\end{tabular}
\vspace{3\baselineskip} \\
\Large\bfseries\titletext\bigskip}
\author{\begin{minipage}[t]{0.8\textwidth}
\normalsize\centering\authortext
\end{minipage}}
\date{}

\begin{document}

\bigskip
\maketitle
\begin{abstract}\normalsize\noindent
\abstracttext
\end{abstract}\normalsize\vspace{\baselineskip}

% \clearpage

% \no indent

\section{Introduction}

The path towards the understanding of the origin of flavour takes us past fermion mass matrices, {which} carry the imprint of the dynamics, if any, determining the structure of fermion masses and mixing. 
Unfortunately, the SM physical flavour parameters, masses and mixings, strictly speaking do not allow to reconstruct the {fermion} mass matrices. {Indeed}, a change of the flavour basis would change the neutrino and charged lepton mass matrices, 
\begin{equation}
\label{eq:basischange}
M_\nu \to V^T_l M_\nu V_l, \qquad 
M_E \to V^T_{e^c} M_E V_l, 
\end{equation}
but not the physical observables. 

A top-bottom perspective is most often taken, plagued however by {a landscape} of equally motivated options for the dynamical origin of the flavour structure (discrete and continuous symmetries~\cite{Frampton:1994rk, Ishimori:2010au, King:2013eh, deMedeirosVarzielas:2005ax,King:2005bj, Berger:2009tt, Adulpravitchai:2009kd}, accidental symmetries~\cite{Wolfenstein:1981kw, Fukuyama:1997ky}, partial compositeness~\cite{Kaplan:1991dc,Redi:2013pga}, extra dimensions~\cite{ArkaniHamed:1998vp,ArkaniHamed:1999dc,Dvali:1999cn,Grossman:1999ra} or even anarchy~\cite{Hall:1999sn} -- see~\cite{Altarelli:2010gt} for an overview and further references), {to be confronted with the limited data available}. Even restricting to flavour symmetries, the large number of possible models reduces the significance of a successful prediction. 

On the other hand, the mere assumption of the existence of a top-bottom perspective allows, as we will see, to infer relevant information on the fermion mass matrices and pursue a bottom-up approach, despite \eq{basischange}. That is because the top-bottom perspective implies the existence of a (unknown) privileged flavour basis, determined by the (unknown) fundamental flavour theory, in which the fermion mass matrices are directly related to the independent fundamental parameters of the theory from which they originate. Because of \eq{basischange}, in all other bases the mass matrix entries will instead be highly correlated, obscure functions of the fundamental parameters. It is the former observation, together with the peculiar experimental values of the flavour parameters (especially their hierarchies), and a simple stability principle~\cite{Marzocca:2014tga} that allows, in some cases, to infer a significant part of the structure of the mass matrices. In turn, this may provide general, model independent hints on the dynamics underlying the structure of the mass matrices.

We extend the work of \cite{Marzocca:2014tga} by applying the stability principle to the small ``solar'' mass squared difference $\Delta m_{12}^2$. This allows us to identify all four stable neutrino mass matrices. 
Interestingly, two of them uniquely correspond to specific neutrino mass patterns. Hence in this context a future determination of the neutrino spectrum will have a chance to uniquely identify \emph{the} neutrino mass texture. We will mainly focus on the case of ``semi-degenerate'' Majorana neutrinos, a neutrino pattern in which two neutrinos are approximately degenerate and the third one is neither degenerate nor hierarchically different, thus implying an overall neutrino mass scale within reach of future experiments. Such a spectrum is uniquely associated to texture A in \Tab{nutextures}. However, the results we will obtain, which include  stringent constraints on the lepton mass and mixing matrices and a mild preference for one of the Majorana phases, will also apply to texture B.

This paper is organized as follows. After introducing the stability principle  in \Sec{stabassumption}, we derive the resulting possible structures of the neutrino mass matrix in \Sec{numass}. Out of these possibilities, we focus on the case of semi-degenerate neutrinos in \Sec{semdeg}, discussing implications for the neutrino and charged lepton contributions to the lepton mixing matrix. In \Sec{othercases} we return to the other possible stable structures for the neutrino mass matrix, before concluding in \Sec{conclusions}. Some details and important proofs are relegated to the appendices: \App{polynomia} shows the derivations of the main results of \Sec{numass} in the limit $\Delta m_{12}^2 \rightarrow 0$. \App{corrections} extends this to finite values of $\Delta m_{12}^2$ for the mass structure discussed in \Sec{semdeg}. \App{finite} gives some details about the definition of the stability principle and finally \App{stabproof} deals with the consequences of the stability assumption on the charged lepton sector.

\section{The stability principle \label{sec:stabassumption}}

The assumption we use is quite basic. We assume that physical quantities, in particular the hierarchical ones (the small ratio of charged fermion masses and the small ratio of the ``solar'' mass squared difference over the ``atmospheric'' one, $|\dm{12}/\dm{23}|$), are stable with respect to small (but finite) variations of the individual matrix entries.\footnote{See also \cite{Peccei:1995fg,Fritzsch:1997fw,Branco:1998bw,Gupta:2016hfm} for alternative approaches to natural mass matrices.} Such an assumption is quite model-independent. It was introduced and systematically used in~\cite{Marzocca:2014tga} in the charged lepton sector. In this work, we extend that study to the entire lepton sector.  

The motivation of the assumption is straight-forward: an ``understanding'' of the smallness of e.g.\ the light fermion masses requires that smallness not to be accidental, i.e.\ to be stable with respect to variation of independent, fundamental parameters. This goes without saying. What we are assuming is that all matrix elements correspond to \emph{independent} fundamental parameters. 

The main caveat to our assumption is then that correlations among different matrix elements might arise, for example as a consequence of a non-abelian symmetry. The latter is of course a concrete possibility, widely studied in the literature. On the other hand, in the light of the fact that experimental hints could have piled up by now in favour of models predicting such correlations, but they have not so far, we do not consider the case in which correlations are absent to be less motivated. Having said that, the principle can be applied (though in a more model-dependent way) to theories predicting correlations among matrix entries as well, by simply expressing the relevant physical quantities in terms of the independent parameters of the theory. 

In the neutrino sector, the stability assumption is most powerful when applied to the solar squared mass difference, as its value is significantly smaller than the atmospheric one, $|\dm{12}/\dm{23}| \approx 0.03 \ll 1$ (for a review on neutrino masses and mixings, including experimental constraints and  details on the notation commonly used, see~\cite{Agashe:2014kda}). As a consequence, $\dm{12}$ is potentially quite sensitive to variations of the neutrino mass matrix entries. Following~\cite{Marzocca:2014tga}, the quantitative formulation of the stability of $\dm{12}$ with respect to variation of a matrix entry $M^\nu_{ij}$ we will use is 
\begin{equation}
\label{eq:stability}
\left|
\frac{\Delta (\dm{12})}{\Delta M^\nu_{ij}} 
\frac{M^\nu_{ij}}{\dm{12}} 
\right| \lesssim 1
\quad
\text{for} 
\quad
|\Delta M^\nu_{ij}| \ll |M^\nu_{ij}| .
\end{equation}
In other words, when $M^\nu_{ij}$ is varied by a small relative amount $|\Delta M^\nu_{ij}/M^\nu_{ij}|$, the corresponding relative variation of $\dm{12}$ should not be much larger, $|\Delta (\dm{12})/\dm{12}| \lesssim |\Delta M^\nu_{ij}/M^\nu_{ij}|$. The definition is of course closely related to the definition of fine-tuning, or sensitivity parameter~\cite{Barbieri:1987fn}, which only differs in the size of the variation, here taken to be small but finite. Such a difference makes our criterium apparently only slightly stronger than the fine-tuning one, but plays an important role, as discussed in \App{finite}. 

\section{Consequences of the stability assumption \label{sec:numass}}

We start from the following proposition about stable Majorana neutrino textures. 
\begin{quote}
In the limit $\dm{12}/\dm{23} \to 0$, the neutrino mass matrix $M_\nu$ satisfies \eq{stability} iff it is in one of the following two forms: 
\begin{equation}
\label{eq:stableforms}
%A: 
\begin{pmatrix}
0 & m & 0 \\
m & 0 & 0 \\
0 & 0 & m_3
\end{pmatrix} %\,,
\qquad
%B:
\text{or}
\qquad
\begin{pmatrix}
0 & m & m' \\
m & 0 & 0 \\
m' & 0 & 0
\end{pmatrix} \,,
\end{equation}
up to a permutation of the rows and columns. 
\end{quote}
{The parameters in \eq{stableforms} can be taken to be real and non-negative without loss of generality.} In order to ensure a non-zero $\dm{23}$, one out of the two parameters in each matrix must be non-zero. On the other hand, one of them can vanish, leading to the four options in \Tab{nutextures}.

\begin{table}
\begin{center}
\begin{tabular}{ccccc}
& A & B & C & D \\
\noalign{\smallskip} 
\hline
\noalign{\medskip} 
&
$
\begin{pmatrix}
 & X &  \\
X &  &  \\ 
 &  & X
\end{pmatrix} 
$
&
$
\begin{pmatrix}
\phantom{X} & X & \phantom{X} \\
X & \phantom{X} &  \\
\phantom{X} & \phantom{X} & 
\end{pmatrix}
$
&
$
\begin{pmatrix}
\phantom{X} & X & X \\
X & \phantom{X} & \phantom{X} \\
X & \phantom{X} & \phantom{X}
\end{pmatrix}
$
&
$
\begin{pmatrix}
\phantom{X} & \phantom{X} & \phantom{X} \\
\phantom{X} & \phantom{X} & \phantom{X} \\
\phantom{X} & \phantom{X} & X
\end{pmatrix}
$
\\ 
\noalign{\medskip} \hline \noalign{\smallskip} 
IH & \cmark & \cmark & \cmark & \xmark
\\
NH & \cmark & \xmark & \xmark & \cmark 
\\
SD & \cmark & \xmark & \xmark & \xmark
\\
\noalign{\smallskip} 
\hline 
\end{tabular}
\end{center}
\caption{The four stable neutrino textures in the $\dm{12}/\dm{23}\to 0$, $\dm{23}\neq 0$ limit, up to permutations of rows and columns. The non-zero entries are denoted by $X$. 
Also shown are the neutrino patterns associated to each texture, inverted hierarchy (IH), normal hierarchy (NH), semi-degeneracy (SD).}
\label{tab:nutextures}
\end{table}

The proof of the proposition makes use of two observations. The first one is that the stability of $\dm{12}$ implies the stability of the parameter
\begin{equation}
\label{eq:delta1}
\Pi \equiv (\dm{12}\dm{23}\dm{13})^2 ,
\end{equation}
i.e.\ it implies,
\begin{equation}
\label{eq:delta2}
\left| \frac{\Delta \Pi}{\Delta M^\nu_{ij}} \frac{M^\nu_{ij}}{\Pi} \right| \lesssim 1,
\end{equation}
as $\dm{23}$, $\dm{13}$ are never very sensitive to variations of the mass matrix entries. The advantage of discussing the stability in terms of $\Pi$ is that $\Pi$ has a calculable polynomial dependence on the matrix entries $M^\nu_{ij}$ and their conjugate $M^{\nu*}_{ij}$. As a consequence, the variation $\Delta \Pi = \Pi(M^\nu_{ij} + \Delta M^\nu_{ij}) - \Pi(M^\nu_{ij})$ that appears in the stability condition is a calculable polynomial in $\Delta M^\nu_{ij}$ and its conjugate. This is shown in \App{polynomia}. 

The second observation is that the stability condition in \eq{delta2} can be re-written as $|(\Pi(M^\nu_{ij} + \Delta M^\nu_{ij}) - \Pi(M^\nu_{ij})) \, M^\nu_{ij}| \lesssim |\Pi \, \Delta M^\nu_{ij}|$, which, in the $\dm{12}\to 0$ limit, becomes
\begin{equation}
\label{eq:stabilitylimit}
M^\nu_{ij} \, \Pi(M^\nu_{ij} + \Delta M^\nu_{ij}) = 0 .
\end{equation}
As $M^\nu_{ij} \, \Pi(M^\nu_{ij} + \Delta M^\nu_{ij})$ is a polynomial in $\Delta M^\nu_{ij}$ and its conjugate, its vanishing for all $\Delta M^\nu_{ij}$ in a neighbourhood of zero (no matter how small) implies the vanishing of all coefficients, in turn polynomials in $M^\nu_{ij}$, $M^{\nu*}_{ij}$. One then obtains simple algebraic conditions on the entries of a stable $M_\nu$, which lead to \eq{stableforms}. This is also shown in \App{polynomia}, where the discussion of the simple $2\times 2$ case can also be found. 

The textures in \Tab{nutextures} are well known and widely studied in the literature, see e.g.~\cite{Petcov:1982ya, Leung:1983ti, Barbieri:1998mq, Altarelli:1999dg, Altarelli:1999wi}. Here we have for the first time rigorously associated them to the stability of the small $\dm{12}$, and shown how they can be obtained from the solution of simple algebraic conditions. Moreover, as we will show in the following, we will obtain relevant information on the size of the entries set to zero in \eq{stableforms}, and as a consequence experimental data to come will provide significant information on the structure of the charged lepton mass matrix as well. 

The textures in \Tab{nutextures} are classified in terms of the neutrino mass pattern they correspond to. Texture D corresponds to normal hierarchy, textures B and C to inverted hierarchy, and texture A can correspond to both, depending on whether the 33 entry is larger or smaller than the 12 entry ($m_3 \gtrless m$ in \eq{stableforms}). Note that it is possible to continuously go from texture A to B and D, and from texture C to B, by making one of the non-zero parameters small. Texture A (if the entries are of the same order of magnitude) corresponds to semi-degenerate neutrinos (see \Sec{semdeg}). Interestingly, future measurements might lead to the unique identification of the neutrino texture. For example, if the sum of neutrino masses turned out to be out of reach and the determination of the sign of $\dm{23}$ pointed at a normal ordering, that would select texture D. If the sum of neutrino masses will end up to be in the range accessible by planned experiments, this will force a semi-degenerate spectrum, and will select texture A. Let us discuss it in greater detail the latter possibility. Most of the analysis in the next section applies to texture B as well. However, experimental data alone does not allow to uniquely identify texture B. This is because the latter corresponds to the same mass pattern as texture C, which however has different implications for the lepton mixing matrices (see \Sec{othercases}).

\section{Semi-degenerate neutrinos (case A) \label{sec:semdeg}}

The case of semi-degenerate neutrinos turns out to be particularly interesting because i) it leads to quite specific forms of the lepton mass matrices and ii) it corresponds to a sum of light neutrino masses $\mtot$ not much below the present experimental limit, perhaps within the reach of possible future generation of experiments aiming at determining the absolute neutrino masses (currently the strongest bound on the absolute neutrino mass scales comes from cosmological probes~\cite{Palanque-Delabrouille:2014jca}, with significant improvements expected from a new generation of spectroscopic surveys and CMB experiments~\cite{Abazajian:2013oma,Basse:2013zua}). As mentioned, most of the results we will obtain, specifically \Sec{nuPMNS}, \Sec{chargedleptons}, \Sec{SU5}, also apply to texture B. 

\subsection{Definition}

As mentioned, we call the light neutrino mass spectrum semi-degenerate\footnote{Sometimes called ``partially degenerate''~\cite{Bilenky:2001rz}, although this terminology is sometimes used with different meanings.} when the two neutrinos $\nu_1$ and $\nu_2$ are quasi-degenerate, and the third neutrino is neither hierarchically larger or smaller than $\nu_{1,2}$, nor degenerate. Semi-degeneracy is compatible with both normal and inverted hierarchy, depending as usual on whether the third neutrino is heavier or lighter than the other two. 

\begin{figure}
 \centering
 \includegraphics[width=\textwidth]{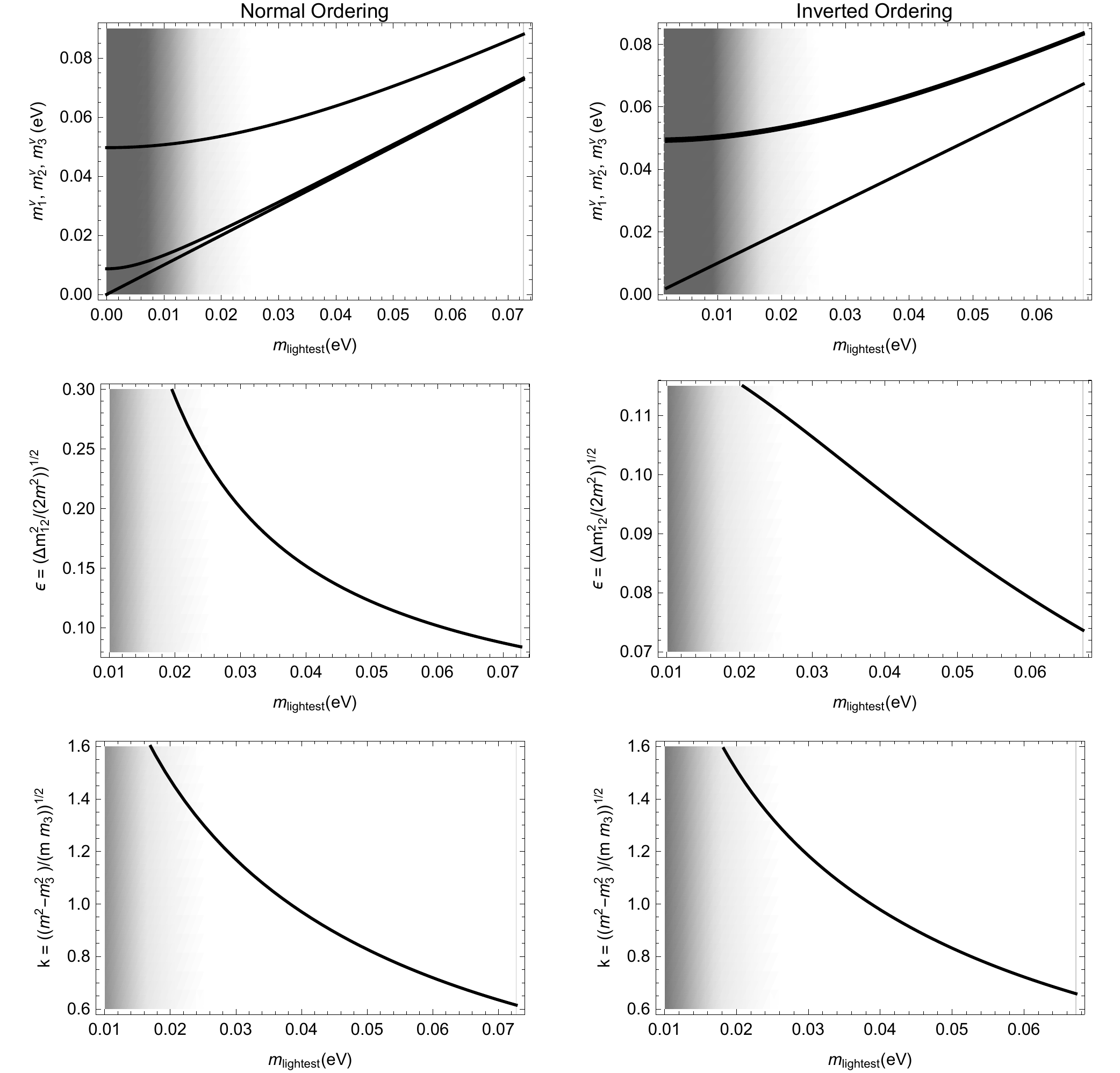}
 \caption{Neutrino masses $m_{i}^\nu$ and the degeneracy parameters as defined in eq.~\eqref{eq:semideg2} in terms of the lightest neutrino mass, for normal and inverted ordering. The vertical line on the right-hand side denotes the current upper bound on the neutrino mass scale~\cite{Agashe:2014kda}, the gray shaded region on the left indicates the violation of the semi-degeneracy condition, i.e.\ $k > {\cal O}(1)$.}
\label{fig:deg}
\end{figure}

\Fig{deg} shows that in a significant range below the present bound on $\mtot$, here taken to be $\mtot < 0.23\eV$~\cite{Agashe:2014kda}, corresponding to the right edge of the plot, the neutrino spectrum is indeed semi-degenerate, with 
\begin{equation}
\label{eq:semideg}
m^2_1\approx m^2_2 \approx m^2\equiv \frac{m^2_1+m^2_2}{2}, 
\qquad
m\sim m_3 ,
\end{equation}
or equivalently
\begin{equation}
\label{eq:semideg2}
\epsilon^2 \equiv \frac{\dm{12}}{2m^2} \ll 1, 
\qquad
k^2\equiv \frac{|m^2 - m_3^2|}{mm_3} =\ord{1}. 
\end{equation}
As a consequence, 
\begin{equation}
\label{eq:PiSD}
\Pi = (\dm{12}\dm{23}\dm{13})^2 \approx (\dm{12})^2(m^2-m_3^2)^4 .
\end{equation}

\subsection{Stability}

Suppose that $\mtot$ is found to lie in the measurable range below the present bound and consequently the light neutrinos are semi-degenerate. Then we know the form of the light neutrino mass matrix, if stable. In the limit in which we neglect $\dm{12}$ correction, it is in the form $A$ in \eq{stableforms}, with $m\sim m_3$, up to permutations of rows and columns. Permutations that can be neglected, as we can always bring $M_\nu$ in the form $A$ of \Tab{nutextures} by properly numbering the three lepton doublets. On top of that, we can use the stability condition to infer the form of $M_\nu$ in the realistic case in which $\dm{12}$ is small but not zero. As shown in \App{corrections}, 
\begin{equation}
\label{eq:SDtexture}
M_\nu = 
\begin{pmatrix}
0 & m & 0 \\
m & 0 & 0 \\
0 & 0 & m_3
\end{pmatrix} + \Delta M_\nu,
\qquad
|\Delta M_\nu| \lesssim 
\begin{pmatrix}
\epsilon^2 m & 0 &  k\epsilon m \\
0 &  \epsilon^2 m &  k\epsilon m \\
 k\epsilon m &  k\epsilon m & 0
\end{pmatrix},
\end{equation}
where $\epsilon$, $k$ are defined in \eq{semideg2}. As discussed in \App{corrections}, similar bounds apply to the case of texture B, in which $m_3 = 0$.

\subsection{Neutrino contribution to the PMNS matrix}
\label{sec:nuPMNS}

The above result determines the natural values of the contribution of the neutrino sector to the PMNS matrix, with significant implications for the structure of the charged lepton sector. A perturbative diagonalization of $M_\nu$ in \eq{SDtexture} yields
\begin{equation}
\label{eq:diag}
M_\nu = U^T_\nu M_\nu^\text{diag} U_\nu ,
\qquad
U_\nu = 
\diag(1,i,1)^*
R_{12}\left(\frac{\pi}{4}-\Delta \right)^{-1}
U' \Psi_\nu ,
\quad
|\Delta| \lesssim \epsilon^2
\end{equation}
where the crucial factor is the 12 rotation $R_{12}$ by an angle that differs from $\pi/4$ by only $\ord{\epsilon^2}$ or less. The factor $i$ is necessary to obtain $M^\text{diag}_\nu>0$, and $\Psi_\nu$ is a diagonal matrix of phases. Finally $U' = \mathbf{1}+\ord{\epsilon}$ is a relatively small correction obtained by combining two unitary transformations in the 13 and 23 blocks. The eigenvalues in $M_\nu^\text{diag}$ are ordered in the standard way. 

\Eq{diag} shows that the diagonalization of the neutrino mass matrix provides an 12 angle very close to $\pi/2$. Therefore, while the neutrino sector provides the leading  contribution to the solar mixing angle $\theta_{12}$, it does not account for the observed deviation of $\theta_{12}$ from $\pi/4$. While the central value of the observed deviation is, according to~\cite{Agashe:2014kda}, $\pi/4 - \theta_{12} \approx 0.2$, \eq{diag} alone would give $\pi/4 - \theta_{12}  = \Delta \lesssim \ord{\epsilon^2}$. \Fig{deg} and the numerical values of $\epsilon^2$ in \Tab{deg} show that this is far from being enough. 

\begin{table}
\renewcommand{\arraystretch}{1.1}
\begin{center}
\begin{tabular}{c|c|c|c|c}
& $\epsilon^2$ & $\epsilon$ & $k$ & FT$_\text{min}$ \\
\hline
NH & 0.0071 & 0.084 & 0.62 & 60 \\
IH & 0.0054 & 0.074 & 0.65 & 80 \\
\end{tabular}
\end{center}
\caption{Numerical values of the quantities defined in \eq{semideg2} in the semi-degenerate regime, for $\mtot = 0.23\eV$, and normal (NH) and inverted (IH) hierarchy. Also shown is the minimal fine-tuning required to obtain a deviation from $\pi/4$ as large as $\pi/4 - \theta_{12}$ in \eq{diag}. }
\label{tab:deg}
\end{table}

\begin{figure}
 \centering
 \includegraphics[width=0.6\textwidth]{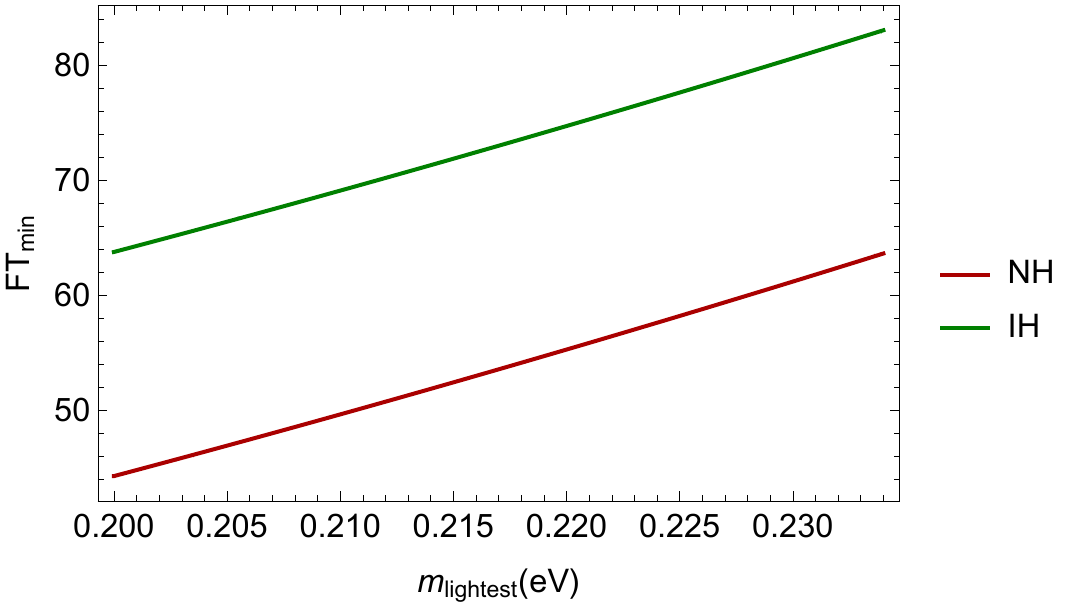}
 \caption{Minimal fine-tuning needed to obtain a deviation from $\pi/4$ as large as $\pi/4 - \theta_{12}$ in \eq{diag}.}
\label{fig:FT}
\end{figure}

We can reverse the argument and estimate how unstable the neutrino mass matrix would be in order for the deviation of the 12 rotation angle to be $\Delta= \pi/4 - \theta_{12}$. For that, it is sufficient to consider the 12 block of $M_\nu$, which, up to a irrelevant constant, is the form
\begin{equation}
\label{eq:12block}
\begin{pmatrix}
a & 1 \\
1 & b
\end{pmatrix},
\end{equation}
with $a$, $b$ complex. The stability of $\dm{12}$ requires $|a|, |b| \lesssim \epsilon^2$. On the other hand, the relation
\begin{equation}
\label{eq:tan2x}
|a| + |b| = 2\frac{|a+b^*|}{|a|-|b|} \tan(2\Delta) \geq 2\tan(2\Delta) \approx 0.8
\end{equation}
forces $|a|+|b|\gg 2 \epsilon^2$. This requires a fine-tuning, because at the same time $|a+b^*|$ needs to be small in order to keep $\epsilon^2 = \dm{12}/(2m^2)$ small, as
\begin{equation}
\label{eq:abstar}
|a+b^*| = \cos(2\Delta) \epsilon^2 \left(1+ \frac{|a|^2+|b|^2}{2} \right) .
\end{equation}
In other words, $a$ and $-b^*$ must be fine-tuned to be approximately the same, with the size of their difference, $|a+b^*|$, much smaller than $|a| \approx |b|\approx (|a| + |b|)/2$. Defining then the fine-tuning to be given by $\FT = [(|a| + |b|)/2]/|a+b^*|$, we have
\begin{equation}
\label{eq:ft}
\FT \geq \frac{\tan(2\Delta)}{\cos(2\Delta)} \frac{2m^2}{\dm{12}} \frac{1}{1+\frac{|a|^2 + |b|^2}{2}} \sim  \frac{\tan(2\Delta)}{\cos(2\Delta)} \frac{2m^2}{\dm{12}} \sim \frac{1}{\epsilon^2} . 
\end{equation}
Numerically, the required minimum fine-tuning turns out to be quite large, as shown in \Tab{deg} and in \Fig{FT}. Strictly speaking, the above formulas hold in the regime $|a|,|b|\lesssim 1$. When $|a|,|b|\gg 1$, the analysis is different, but the outcome is similar. 

In summary, the neutrino contribution to the solar mixing angle is expected to be very close to $\pi/4$. The deviation of $\theta_{12}$ from $\pi/4$, as well as the large value of the atmospheric angle $\theta_{23}$, must therefore originate in the charged lepton sector. This has strong implications on its structure, as we will see now.

\subsection{Charged leptons}
\label{sec:chargedleptons}

As we have seen in the previous subsection, a semi-degenerate neutrino spectrum requires the measured deviation of $\theta_{12}$ from $\pi/4$ and $\theta_{23}$ to originate mostly from the charged lepton sector. Following~\cite{Marzocca:2014tga}, in this subsection we show that i) this is possible and compatible with the stability of the charged lepton sector, ii) the deviation of $\theta_{12}$ from $\pi/4$ and the size $\theta_{13}$ turn out to be essentially independent, and iii) the charged lepton mass matrix needs to take quite a specific form. Our approach therefore provides several pieces of the lepton flavour puzzle, as the neutrino and charged lepton mass matrices are a direct emanation of the physics from which lepton flavour originates. 

The charged lepton mass matrix, in particular its last and leading row, can be reconstructed from
\begin{equation}
\label{eq:reconstruction}
M_E = U^T_{e^c} M^\text{diag}_E U_e ,
\end{equation}
as $M^\text{diag}_E = \diag(m_e,m_\mu,m_\tau)$ is known, $U_e$ can be obtained from $U_e = U U_\nu$, with $U$ denoting the PMNS matrix. Here $U$ and $U_\nu$ are now known (up to phases) from data and \eq{diag} respectively, and $U_{e^c}$ turns out to be constrained by stability. In order to reconstruct $M_E$ from \eq{reconstruction}, let us start with obtaining $U_e$. 

\subsubsection{$U_e$}

\begin{figure}
\begin{center}
%
%\vspace*{-1.5cm}
\fbox{\footnotesize Normal Ordering} \\[0.2cm]
\vspace*{-0.2cm}
\hspace*{-0.65cm} 
\begin{minipage}{0.5\linewidth}
\begin{center}
	\hspace{0.75cm} \mbox{\footnotesize (a)} \\[0.5mm]
	\includegraphics[width=70mm]{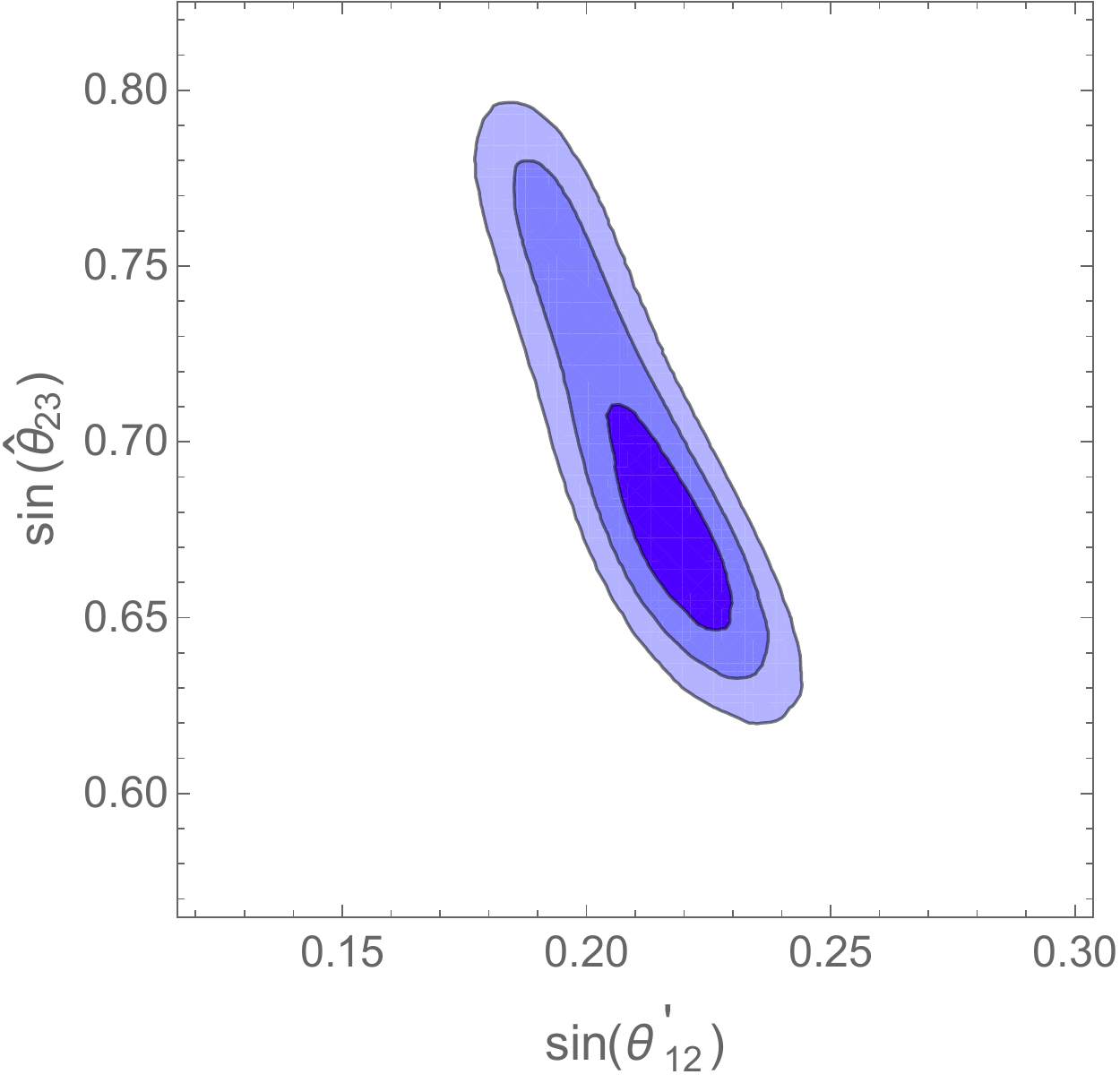}
\end{center}
\end{minipage}
\begin{minipage}{0.5\linewidth}
\begin{center}
	\hspace{0.5cm} \mbox{\footnotesize (b)} \\[0.5mm]
	\includegraphics[width=70mm]{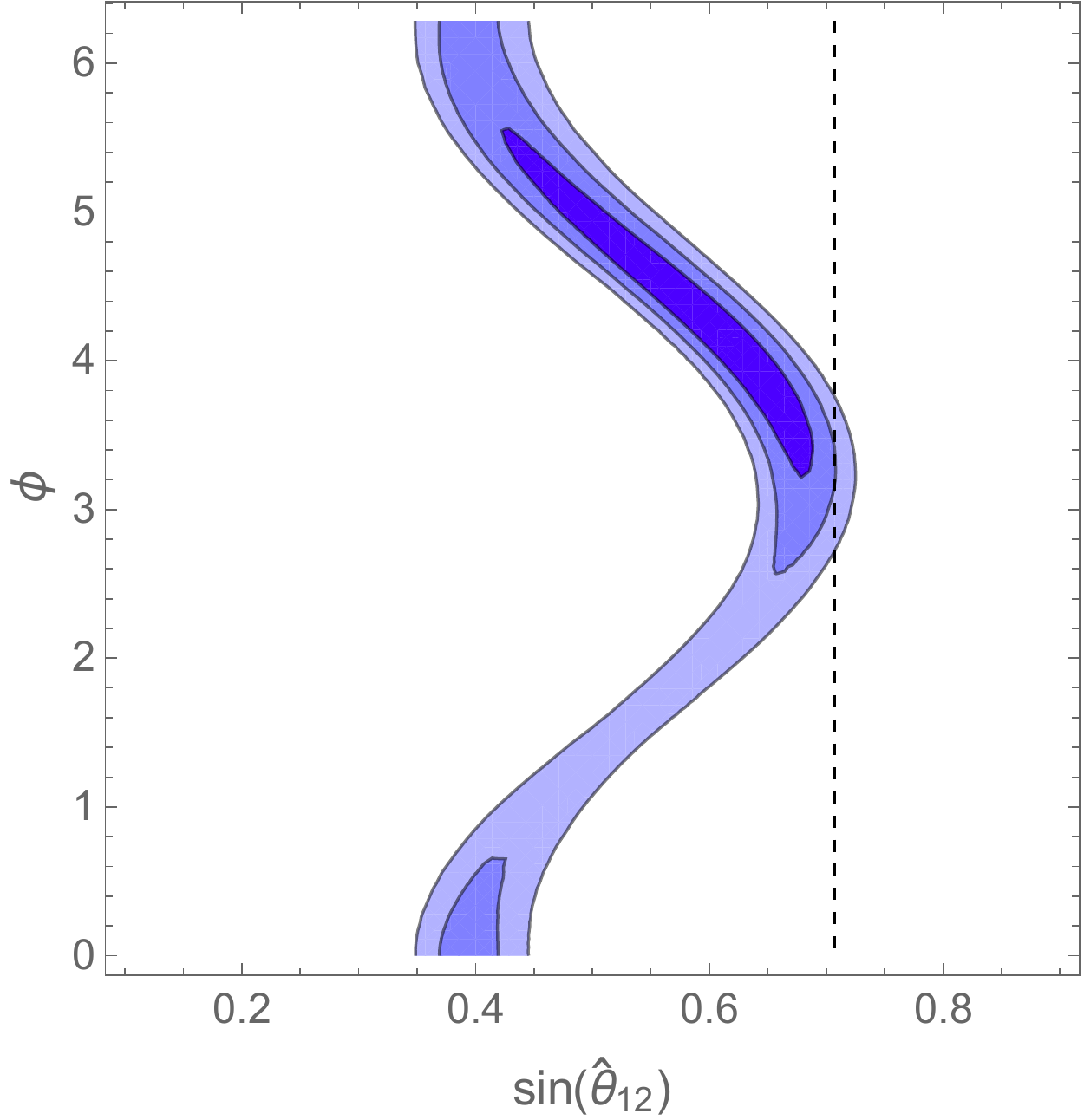}
	\end{center}
\end{minipage} 
\\
\fbox{\footnotesize Inverted Ordering} \\[0.2cm]
\vspace*{-0.2cm}
\hspace*{-0.65cm} 
\begin{minipage}{0.5\linewidth}
\begin{center}
	\hspace{0.75cm} \mbox{\footnotesize (c)} \\[0.5mm]
	\includegraphics[width=70mm]{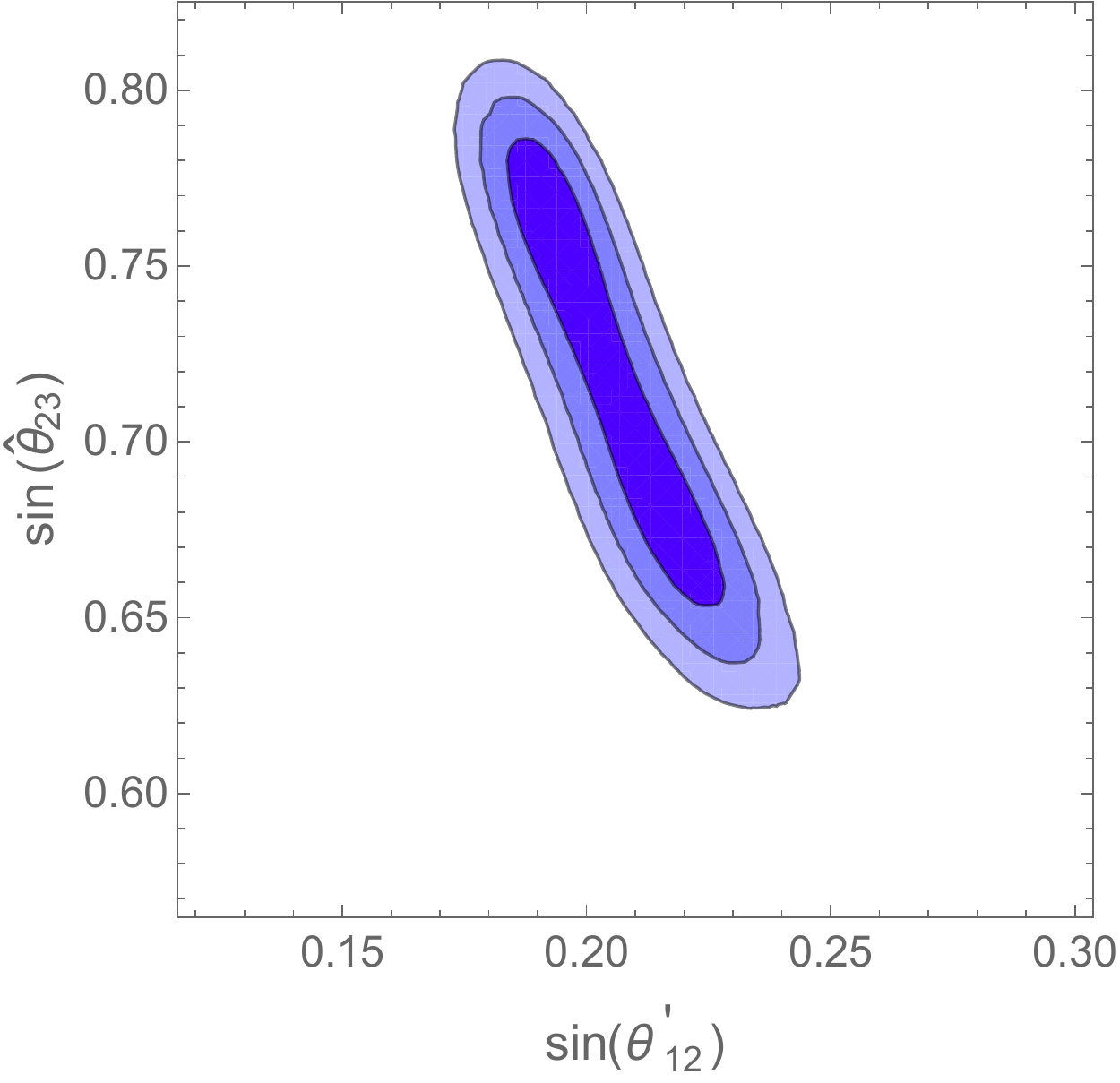}\
	\end{center}
\end{minipage}
\begin{minipage}{0.5\linewidth}
\begin{center}
	\hspace{0.5cm} \mbox{\footnotesize (d)} \\[0.5mm]
	\includegraphics[width=70mm]{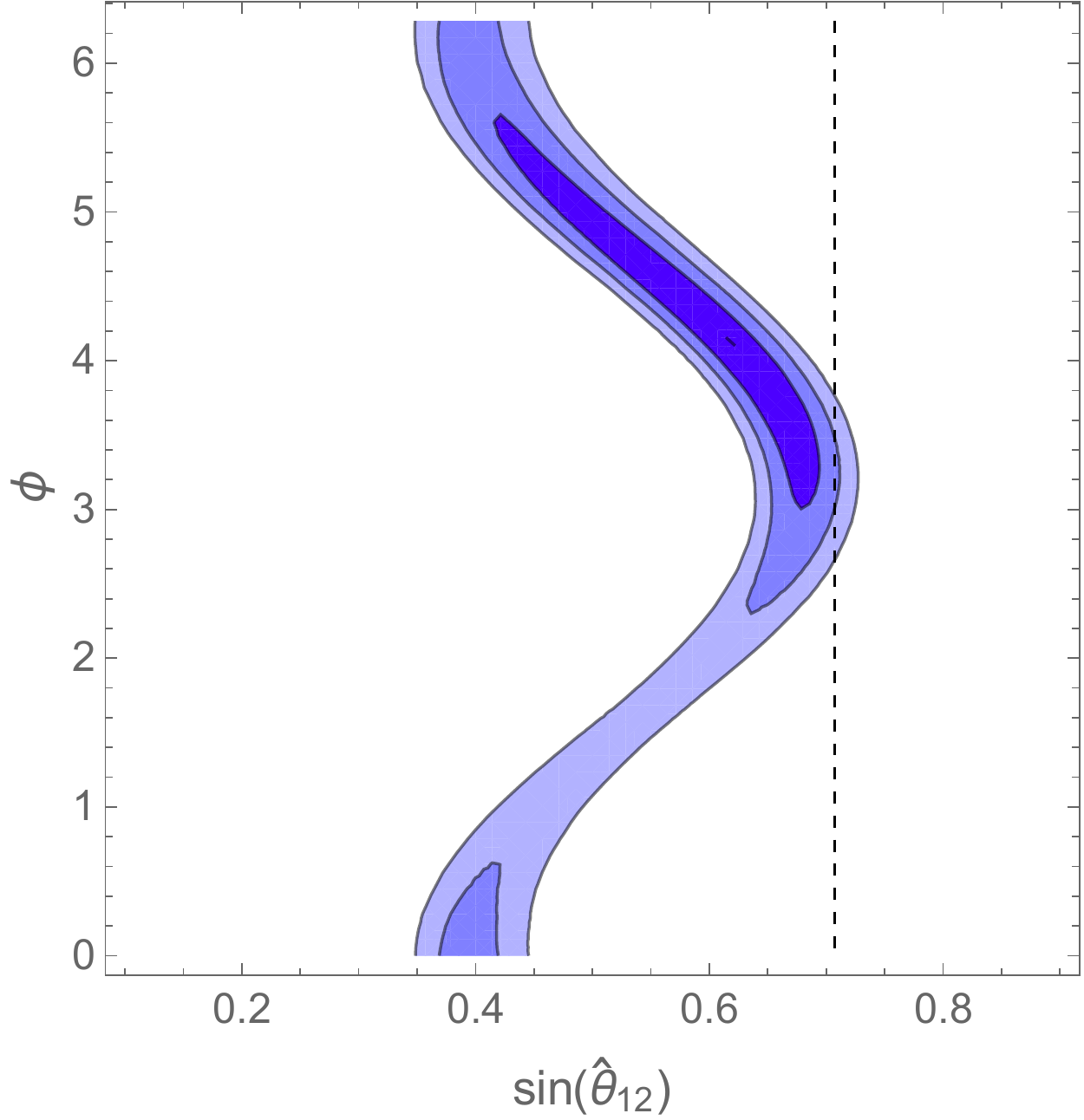}
	\end{center}
\end{minipage} 
 \end{center}
\vspace*{-0.5cm}
\caption{\label{fig:fit} 
\small  $68\%$, $95\%$, and $99.7\%$ confidence level contours 
in the $(\sin \theta_{12}^\prime, \sin\hat \theta_{23})$ (a,c) and $(\sin \hat\theta_{12}, \phi)$ (b,d) planes. We construct the likelihood function using the results of the recent global fit of neutrino oscillation data from ref.~\cite{Capozzi:2016rtj} for normal ordering (upper row) and inverted ordering (lower row) of neutrino masses. In plots (a,c) we use only the constraints on $\sin^2 \theta_{13}$ and $\sin^2 \theta_{23}$. In plots (b,d) we include also the constraints on $\sin^2 \theta_{12}$ and $\delta$, and we marginalize over $\sin \theta_{12}^\prime$ and $\sin \hat \theta_{23}$. The dashed line indicates a value of $\hat \theta_{12} = \pi/4$.
}
\end{figure}

In order to obtain $U_e$ from $U_e = U U_\nu$, it is convenient to write the PMNS matrix $U$ using the parameterisation in~\cite{Fritzsch:1997fw,Marzocca:2011dh} ({see also~\cite{Fritzsch:1997st}}). 
\begin{equation}
\label{eq:para}
U = \hat\Phi_e R_{12}(\theta'_{12}) 
\begin{pmatrix}
1 & & \\
& e^{-i\phi} & \\
& & 1
\end{pmatrix}
R_{23}(\hat\theta_{23}) R_{12}(\hat\theta_{12})\hat\Phi_\nu ,
\end{equation}
where {$\hat\Phi_\nu = \diag(1,e^{i\hat\alpha}, e^{i\hat\beta})$} contains the Majorana phases and $\hat\Phi_e$ is an unphysical diagonal matrices of phases. In this parameterisation, $\hat\theta_{23}$, $\hat\theta_{12}$ are close to the standard PMNS parameters $\theta_{23}$, $\theta_{12}$~\cite{Agashe:2014kda} respectively, while $\theta_{12}'$ mainly determines the $\theta_{13}$ angle (and $\phi$ the CP-violating phase $\delta$):
\begin{align}
\label{eq:paraconversion}
\tan\theta_{12} &= \tan\hat\theta_{12} \left| \frac{1+e^{-i\phi} \tan\theta_{12}'\cos\hat\theta_{23}/\tan \hat \theta_{12}}{1-e^{i\phi} \tan\theta_{12}'\tan\hat\theta_{12} \cos\hat\theta_{23}} \right| \hspace{-2.5cm}  &
\sin\delta = \sin\phi \, \frac{\sin2\hat\theta_{12}}{\sin2\theta_{12}}  & \,,
	\nonumber \\[1.5mm]
\sin \theta_{13} &= \sin \theta'_{12} \sin \hat\theta_{23} &
e^{i(\alpha-\beta)} = e^{i(\hat\alpha-\hat\beta)} \, \text{Ph}(1- \tan\theta'_{12} \tan\hat \theta_{12} \cos\hat\theta_{23} e^{i\phi}) & \,,
	 \\[4mm] 
\tan\theta_{23} &= \tan \hat\theta_{23} \cos\theta_{12}'  &
e^{i\beta} = e^{i\hat\beta} \, \text{Ph}(1 + \tan\theta'_{12}/\tan\hat\theta_{12} \cos\hat\theta_{23} e^{-i\phi}) & \,.
	\nonumber
\end{align} 
\Eq{paraconversion} also shows that the the ``Dirac'' phases $\phi$ and $\delta$, as well as the Majorana phases $\hat\alpha$, $\hat\beta$ and the corresponding ones in the standard parameterisation, $\alpha$, $\beta$~\cite{Feruglio:2002af}, are also relatively close.
A numerical fit of the parameters $\hat\theta_{23},\hat\theta_{12},\theta'_{12},\phi$ based on the updated constraints in~\cite{Capozzi:2016rtj} is shown in \Fig{fit}.\footnote{Our $\phi$ differs from that of~\cite{Marzocca:2013cr} by a sign. \Eqs{paraconversion} determine $\delta$ up to a twofold ambiguity. A full formula is
\begin{equation*}
\label{eq:footeq}
e^{i\delta} \tan\theta_{12} = e^{i\phi} \frac{\tan\hat\theta_{12} + e^{-i\phi} \tan\theta'_{12} \cos\hat\theta_{23} }{1- e^{i\phi} \tan\hat\theta_{12} \tan\theta'_{12} \cos\hat\theta_{23}} .
\end{equation*}}

%%%%%%%%%%%%%%%%%%%%%%%%%%%%%%%%%%%%%%%

We can now combine $U$ and $U_\nu$ to obtain $U_e$. We will neglect here the $\lesssim \epsilon$ contributions from $U'$ in \eq{diag}, as these turn out to be subdominant in the vast part of the parameter space (a notable exception is when these contributions saturate the naturalness bound and cancel the PMNS contribution to $\theta^e_{12}$, in which case some of the bounds quoted in this section can be avoided, see \App{stabproof} for details). 
 $U_e$ turns out to be in the same form as $U$, 
\begin{equation}
\label{eq:Ue}
U_e = \Phi_e R_{12}(\theta'_{12}) 
\begin{pmatrix}
1 & & \\
& e^{-i\phi_e} & \\
& & 1
\end{pmatrix}
R_{23}(\hat\theta_{23}) R_{12}(\theta^e_{12}) \Psi_e ,
\end{equation}
where $\Phi_e$, $\Psi_e$ are again irrelevant diagonal matrices of phases. Note that $U_e$ is determined by the PMNS angles $\theta_{12}'$ and $\hat\theta_{23}$, and by the angle $\theta^e_{12}$, which is the result of combining the PMNS rotation $\hat\theta_{12}$ and the $\pi/4$ neutrino rotation. In the absence of phases, we would have $\theta_e = \pi/4 \pm \hat\theta_{12}$. Because the combination of the two rotations does involve phases, we have instead
\begin{equation}
\label{eq:thetaerange}
\frac{\pi}{4}-\hat\theta_{12} \leq \theta^e_{12} \leq \frac{\pi}{4}+\hat\theta_{12} .
\end{equation}
The precise value of $\theta^e_{12}$ in the above interval (and the phase $\phi_e$) is known if the Dirac and the Majorana phases ($\phi$, $\hat\alpha$, $\hat\beta$) are, 

\begin{equation}
\label{eq:thetae}
\tan\theta_{12}^e  = \fracwithdelims{|}{|}{1-e^{i\xi} \tan\hat\theta_{12} }{1+e^{-i\xi} \tan\hat\theta_{12}} ,
\qquad
e^{i\xi} = e^{i(\hat\alpha-\pi/2)} .
\end{equation}
For completeness, the phase $\phi_e$ is given by
\begin{equation}
e^{i\phi_e} = - e^{i(\phi-\xi)} \, \text{Ph}\left(\frac{1 - e^{i\xi} \tan\hat\theta_{12} }{1 + e^{-i\xi} \tan\hat\theta_{12}}\right) .
\end{equation}

From the phenomenological point of view, it is important to note that there is at present a $2\sigma$ preference for $\hat\theta_{12}$ to be different from $\pi/4$ (corresponding to the vertical dashed lines in plots (b,d)), which implies $\theta^e_{12} \neq 0, \pi/2$ for any value of the phases. In order to strengthen this conclusion, a better experimental determination of $\hat\theta_{12}$, i.e.\ $|U_{32}/U_{31}|$, or $\delta$ in the standard parameterisation, is needed. 

From a model-building point of view, a relevant remark concerns the expected size of $\theta^e_{12}$. While in principle $\theta^e_{12}$ can be anywhere in the range in \eq{thetae}, we argue that simple naturalness considerations mildly favour the lower end of that interval, which in turn has implications for the value of the Majorana phases. Let us first remind that $\hat\theta_{12}$, and the PMNS matrix in general, is a derived quantity, obtained by combining the charged lepton and neutrino rotations $\theta^e_{12}$ and $\pi/4$, directly related to the underlying mass matrices. What we are doing here is inverting that relation and reconstructing $\theta^e_{12}$ in terms of $\hat\theta_{12}$ and $\pi/4$. Now, $\hat\theta_{12}$ is relatively close to $\pi/4$, the neutrino contribution to it. If $\hat\theta_{12}$ turned out to be very close to $\pi/4$, this would suggest that the charged lepton correction $\theta^e_{12}$ to $\pi/4$ is small, $\theta^e_{12} \approx 0$\footnote{Or $\theta^e_{12} \approx \pi/2$. The two cases are however equivalent, as an exchange of the labeling of the first two lepton doublets, $l_1 \leftrightarrow l_2$, shows.} (although, by finely adjusting phases, $\hat\theta_{12}\approx \pi/4$ could also be obtained for $\theta^e_{12} = \pi/4$) and $e^{i\xi} \approx  1$. The present experimental information suggests that $\hat\theta_{12}$ is relatively close to $\pi/4$, but not extremely close. Still, such a closeness might suggest that $\theta^e_{12}$ lies near the lower bound of the interval in \eq{thetaerange} and that $\hat\alpha = \ord{\pi/2}$. Again, a better experimental determination of $\hat\theta_{12}$ would be welcome to assess the size of $\pi/4 - \hat\theta_{12}$. 

\begin{figure}
 \centering
 \includegraphics[width=0.7\textwidth]{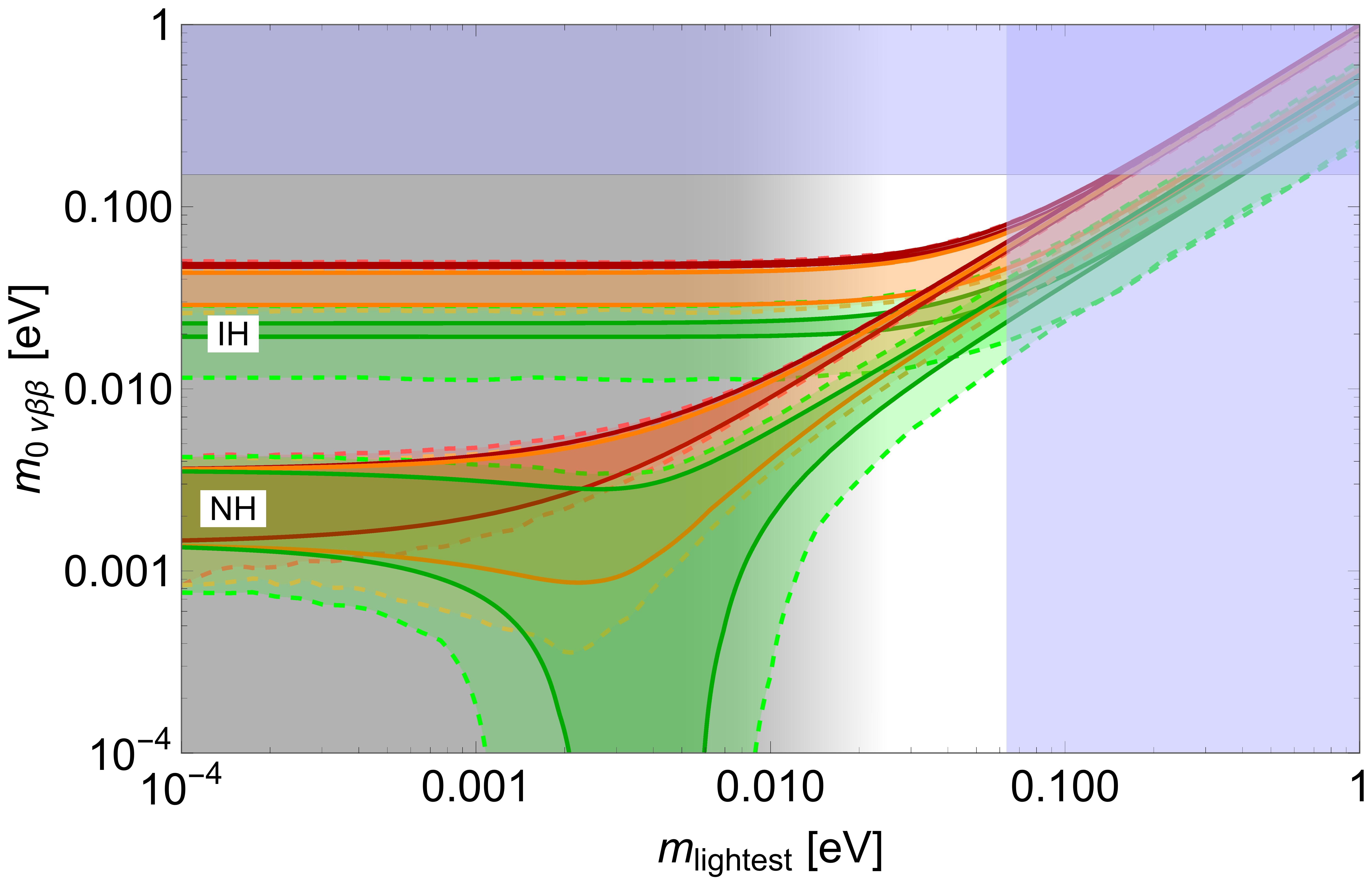}
 \caption{{Value of} $m_{0 \nu \beta \beta}$ for different values of the PMNS Majorana phase $\hat \alpha$: $\hat \alpha = \pi/2$ (green, preferred value),  $\hat \alpha = \pi/4$ (orange), $\hat \alpha = 0$ (red). The blue shaded regions on the top and righthand side of the plot denote current experimental bounds, the grey shaded region indicates the mass range disfavoured by the requirement of semi-degeneracy, see also Fig.~\ref{fig:deg}.}
\label{fig:m0bb}
\end{figure}

The relation $e^{i\xi} \approx  1$, if taken seriously, would lead to a prediction for one of the Majorana phases, $\hat\alpha \approx \pi/2$. This in turn would have interesting consequences for the mass parameter of neutrinoless double-$\beta$ decay $m_{0\nu\beta\beta} \equiv |\sum U_{ei}^2 m_i|$. In the semi-degenerate regime, neglecting $\ord{\theta^2_{13}}$ effects, and approximating $\alpha \approx \hat\alpha$, the only phase entering $m_{0\nu\beta\beta}$ is $\hat\alpha$,
\begin{equation}
m_{0\nu\beta\beta} \approx 
m\, |\cos^2\theta_{12} + e^{2 i \hat\alpha } \sin^2\theta_{12} | .
\label{eq:0n2ba}
\end{equation} 
In the semi-degenerate regime, one has therefore an experimentally accessible value of $m_{0\nu\beta\beta} = \ord{m}$, but the $\hat\alpha \approx \pi/2$ relation implies a partial cancellation between the first two terms in $m_{0\nu\beta\beta}$, forcing this parameter towards the lower edge of the allowed band
\begin{equation}
m_{0\nu\beta\beta} \approx m \cos2\theta_{12} .
\label{eq:0n2b2}
\end{equation}
This is demonstrated in Fig.~\ref{fig:m0bb}, where we show the predictions for $m_{0\nu\beta\beta}$ in terms of the lightest neutrino mass for both normal and inverted hierarchy. The color-coding refers to different values of $\hat \alpha = \{0, \pi/4, \pi/2 \}$, where the green band denotes our preferred value of $\hat \alpha = \pi/2$. The solid lines correspond to fixing the mixing angles and mass splittings to their best-fit values according to \cite{Capozzi:2016rtj} while varying the three phases $\alpha,\beta,\delta$ in the PMNS matrix (subject to the constraint on $\hat \alpha$). The dashed lines refer to the 3$\sigma$ contour, where we have constructed the $\chi^2$ function based on the distributions shown in \cite{Capozzi:2016rtj} for $\theta_{12}, \theta_{13}, \theta_{23}, \delta, \dm{12}$ and $\dm{23}$. As in Fig.~\ref{fig:fit}, we neglect any cross-correlations between these parameters. We note that restricting the Majorana phase $\hat \alpha$, even while all the other phases are unconstrained, significantly reduces the uncertainty on $m_{0\nu\beta\beta}$.  In addition, the blue shaded regions in Fig.~\ref{fig:m0bb} denote the current 3$\sigma$ bounds on $m_{0 \nu \beta \beta}$~\cite{PhysRevLett.110.062502} and on the sum of neutrino masses as constrained by cosmological probes~\cite{Palanque-Delabrouille:2014jca}, respectively (see \cite{Dell'Oro:2016dbc} for a recent comprehensive review). The grey shaded region on the lefthand side indicates the region disfavoured by the requirement of semi-degeneracy, $k \gtrsim {\cal O}(1)$. The remaining allowed window will be probed  in a variety of future experiments: (near) future neutrinoless double-$\beta$ decay experiments are expected to reach a sensitivity for $m_{0 \nu \beta \beta}$ of ${\cal O}(0.1~\text{eV})$ or possibly even ${\cal O}(0.01~\text{eV})$~\cite{Dell'Oro:2016dbc} while cosmological bounds on the sum of neutrino masses are expected to improve with future CMB missions and with upcoming spectroscopic surveys (such as BOSS, DESI and EUCLID), reducing the $1\sigma$ uncertainty on $\mtot$ to ${\cal O}(10 \text{ meV})$~\cite{Abazajian:2013oma,Basse:2013zua}. It should be noted that cosmological bounds on the neutrino mass mentioned above are based on the assumption of $\Lambda$CDM cosmology, whereas the
 tritium decay experiment KATRIN~\cite{Osipowicz:2001sq} is expected to lower the bound on the absolute neutrino mass under laboratory conditions from the current $\sim 2$~eV~\cite{Kraus:2004zw, Aseev:2011dq} to about 0.35~eV.

\subsubsection{$M_E$}

We can now reconstruct the charged lepton mass matrix, in particular its leading rows, from $M_E = U^T_{e^c} M^\text{diag}_E U_e$. The previous formula and what we have learned about $U_e$ force at least two (more likely three) large $\ord{m_\tau}$ entries in the last row. Under this condition, the stability constraints on the charged lepton mass force the third row of $|M_E|$ to be fully determined by the PMNS parameters, up to corrections of relative order $\ord{m_\mu^2/m_\tau^2} = \ord{0.003}$~\cite{Agashe:2014kda},
\begin{equation}
\label{eq:thirdrow}
|M_{3i}^E| \approx |U^e_{3i}| m_\tau = (s^e_{12} \hat s_{23} , c^e_{12} \hat s_{23} , \hat c_{23}) \, m_\tau ,
\end{equation}
where $\theta^e_{12}$ is related to the PMNS parameters by \eq{thetae} and the ranges of the PMNS parameters $\hat\theta_{12}$, $\hat\theta_{23}$ are shown in \Fig{fit}. The range in \eq{thetaerange}, assuming without loss of generality $\tan\theta^e_{12} \leq 1$, and using the present central values of the PMNS parameters, becomes
\begin{equation}
\label{eq:thetaerangenum}
0.13 \leq \tan\theta^e_{12} \leq 1 ,
\end{equation}
with $\tan\theta^e_{12} = 0$ disfavoured at $2\sigma$ (and a possible preference for values around the lower bound from the naturalness considerations in the previous subsection). This means that $M^E_{31} = 0$ is also disfavoured and $|M^E_{31}| \gtrsim 0.13 \, m_\tau$ is preferred. Note that this preference for tan$\theta_{12}^e \neq 0$ may however fade away for specific values of the $\lesssim \epsilon$ contributions to lepton mixing from $U'$ in \eq{diag}, cf.\ \App{stabproof}.

The stability of the charged lepton masses also provides information on the first two rows of $M_E$. Denoting $t_e \equiv \tan\theta^e_{12}$ and $t' \equiv \tan\theta'_{12}$, we can show that there exists a $t$, with
\begin{equation}
t' \leq t \leq \max(t',t_e) ,
\label{eq:trange}
\end{equation}
such that
\begin{equation}
\label{eq:ME1}
\renewcommand{\arraystretch}{1.5}
|M_E| = 
\begin{pmatrix}
\lesssim m_e & \displaystyle \lesssim m_e \min\left(\frac{1}{t},\frac{1}{t_e}\right) & \displaystyle \lesssim \frac{m_e}{t} \\[3mm]
\lesssim t \, m_\mu & \displaystyle \lesssim m_\mu \min\left(1,\frac{t}{t_e}\right) & \sim m_\mu \\[3.5mm]
\sim t_e m_\tau & \sim m_\tau & \sim m_\tau
\end{pmatrix} P_{23}, \quad 
\begin{gathered}
t' \leq t \leq  \max(t',t_e) \\
0.13 \lesssim t_e \leq 1 \\ 
t' \approx 0.22 
\end{gathered} ,
\end{equation}
where we have used our best fits for $t' = \tan\theta'_{12}$ and for the lower bound of $t_e$. $P_{23}$ represents a permutation matrix that is either the identity or exchanges the last two columns. Note that the above results improve on those in~\cite{Marzocca:2014tga}, where the range of $t$ (there called $1/k$) was looser and the constraints on $M_E$ milder. As a byproduct, we also obtain stability bounds on $U_{e^c}$,  
\begin{equation}
\renewcommand{\arraystretch}{1.8}
|U_{e^c}| \approx 
\begin{pmatrix}
1 & 
\displaystyle \lesssim \frac{m_e}{m_\mu t} & 
\displaystyle \lesssim \frac{m_e}{m_\tau t} \\
\displaystyle \lesssim \frac{m_e}{m_\mu t} & 
1 & 
\displaystyle \lesssim \frac{m_\mu}{m_\tau} \\
\displaystyle \lesssim \frac{m_e}{m_\tau t} &
\displaystyle \lesssim \frac{m_\mu}{m_\tau} &
1
\end{pmatrix},
\label{eq:Ucstability}
\end{equation}
which will be used in the next subsection. Eqs.~(\ref{eq:ME1},\ref{eq:Ucstability}) are proven in \App{stabproof}.

In summary, with no theoretical assumption but the stability of the small $\dm{12}$ squared mass difference and of the electron and muon mass, data leads us in the case of semi-degenerate neutrinos to a unique leading order texture for the charged lepton mass matrices
\begin{equation}
\label{eq:semitexture}
M_\nu = 
\begin{pmatrix}
& X & \\
X & & \\
& & X
\end{pmatrix} + \text{smaller}, 
\qquad
M_E = 
\begin{pmatrix}
&&\\
&&\\
X? & X & X
\end{pmatrix}  + \text{smaller}, 
\end{equation}
which represents a model-independent handle on the origin of lepton flavour. One can for example ask in full generality the question whether the above texture, in the limit of vanishing corrections, can be obtained from the symmetric limit of a generic flavour symmetry acting (possibly independently) on the lepton fields. It is not difficult to show that this is not the case~\cite{YR}. 

\subsection{Compatibility with SU(5)}
\label{sec:SU5}

In SU(5), the matrix $U_{e^c}$ is related to $U^T_d$, where $U_d$ is the down quark contribution to the CKM matrix $V$, $V  = U_u U^\dagger_d$. In the unbroken SU(5) limit, $U_{e^c} = U^T_d$, but SU(5) breaking effects can introduce differences, governed by SU(5) Clebsh factors, typically of order one~\cite{Marzocca:2011dh}. Stability in the quark sector suggests that $V \approx U^\dagger_d$, and the absence of cancellations in the determination of the CKM matrix requires that the $U_d$ angles are not  much larger than the CKM ones. 

Let us consider the case in which $V \approx U^\dagger_d$. We can then compare $|U^d_{12}| \approx \sin\theta_C \approx 0.22$, where $\theta_C$ is the Cabibbo angle, with its SU(5) counterpart $|U^{e^c}_{21}| \lesssim m_e/m_\mu/t \leq m_e/m_\mu/t' \sim 0.02$. Clearly, an SU(5) realisation of the stable semi-degeneracy textures studied in the previous subsections requires quite important Clebsch factors. The simplest possibility is the following
\begin{equation}
M_E = 
\begin{pmatrix}
C_1 \tau & C_2 & C_3 \\
B_1 \tau & B_2 & B_3 \\
A_1 \tau & A_2 & A_3
\end{pmatrix},
\qquad
M^T_D = 
\begin{pmatrix}
3 C_1 \tau & 3 C_2 & 3 C_3 \\
B_1 \tau/3 & B_2/3 & B_3/3 \\
A_1 \tau & A_2 & A_3
\end{pmatrix},
\label{eq:SU5}
\end{equation}
with $A_i \gg B_i \gg C_i$ and $\tau \sim t' \approx 0.2$. Such textures give in first approximation $m_\tau \approx m_b$, $m_\mu \approx 3 m_s$, $m_e \approx m_d/3$ at the unification scale, in reasonable agreement with data, and $|U^{e^c}_{21}| \approx B/C \approx \sin\theta_C / 9$, in agreement with the numerical figures above. The fact that $t'$ happens to be close to the Cabibbo angle implies that $|M^D_{12}| \approx |M^D_{21}|$ in \eq{SU5}.

\section{Hierarchical neutrinos (cases B, C, D) \label{sec:othercases}}

Let us now consider the situation when we drop the requirement of a semi-degenerate neutrino mass spectrum, i.e.\ cases B, C and D in \Tab{nutextures}. As discussed, case B gives the same results for $U_\nu$, $U_e$, and $M_E$ as case A. In case C, the corresponding neutrino mixing matrix $U_\nu$ is at leading order in $\epsilon$ given by
\begin{equation}
U_\nu = \text{Diag}(1,i,1) R_{12}(\pi/4) R_{23} (\theta^B_{23}) \Phi_B\,,
\end{equation}
with $\sin(\theta_{23}^B) = 1/\sqrt{1 + |M^\nu_{12}/M^\nu_{13}|^2}$ and $\Phi_B$ a diagonal matrix of phases. Constructing $U_e = U U_\nu$, we find that, contrary to the semi-degenerate case discussed above, both the $U_{13}^e$ and $U_{31}^e$ elements are no longer bounded from below. This significantly weakens the constraints arising from the charged lepton sector, in fact this is just the situation discussed in \App{stabproof} in the case that the ${\cal O}(\epsilon)$ corrections in $U'$ of Eq.~\eqref{eq:diag} cancel the ${\cal O}(t',t_e)$ contributions in $U_{31,13}^e$, cf.\ Eqs.~\eqref{eq:U31e} and \eqref{eq:U13e}. The constraints on $M_E$ and $U_{e^c}$ then relax to Eqs.~\eqref{eq:MEepsilon} and \eqref{eq:Uecepsilon}.

Finally in case D, the leading order contribution to $U_\nu$ is a rotation in the 12-block, whose size (set by the subleading contributions to $M_\nu$) is a free parameter. Constructing $U_e = U U_\nu$ and comparing to the semi-degenerate case A, this implies that the parameter $\theta_{12}^e$ is now no longer constrained. This turns out to only mildly weaken the bounds on the charged lepton sector.

\section{Conclusions\label{sec:conclusions}}

We considered a bottom-up approach to lepton flavour based on a simple and motivated hypothesis, the stability of (small) physical quantities with respect to the variations of individual matrix elements, assumed to correspond to independent parameters of an underlying flavour theory. 

The technical tools gathered in the Appendices allow to translate such an hypothesis into a set of algebraic conditions on the matrix entries. When applied to the stability of the small solar squared mass difference $\dm{12}$, those conditions identify, at the leading order in $\dm{12}$, a set of only four possible stable textures for the neutrino mass matrix, see Tab.~\ref{tab:nutextures}. While those textures have been previously studied, we have for the first time rigorously associated them to the stability of $\dm{12}$, and obtained them from the solution of simple algebraic conditions. More important, the stability hypothesis allows to set bounds on the size of the subleading entries, and to draw consequences for the structure of the charged lepton mass matrix. 

The four textures are characterised by their neutrino mass pattern. Interestingly, two of them correspond to a specific mass pattern: texture D corresponds to a third neutrino hierarchically heavier than the other two and texture A corresponds to what we call a semi-degenerate neutrino spectrum, i.e.\ to two quasi degenerate neutrinos and a third neutrino neither hierarchically larger nor smaller than the other two (a spectrum compatible with both normal and inverted ordering, depending on whether the third neutrino is heavier or lighter). Therefore determining the neutrino spectrum might allow to uniquely identify the neutrino mass texture. The semi-degenerate pattern is particularly interesting both from the experimental and theoretical points of view: it corresponds to an overall neutrino mass scale not much below the present experimental limit, perhaps within the reach of future experiments aiming at determining the absolute neutrino mass scale; and it leads, under the stability hypothesis, to quite a specific form of both the neutrino and charged lepton mass matrices. We therefore mostly concentrated on the semi-degenerate case. However, most of the results we obtained also hold in the case of texture B. 

The neutrino contribution $U_\nu$ to the PMNS matrix $U = U_e U^\dagger_\nu$ is then precisely predicted, up to phase rotations, by the stability condition. In particular, $U_\nu$ provides an almost maximal contribution to the solar mixing angle, with a deviation predicted by stability to be $\lesssim 0.01$. The latter can hardly account for the deviation from maximal of the solar mixing angle, $\pi/4 - \theta_{12} \approx 0.2$, unless a $\mathcal{O}(50$--$100)$ fine-tuning is accepted. 

With $U_\nu$ determined by stability and $U$ largely known from the experiment, the left-handed charged lepton contribution to the PMNS matrix $U_e$ can be reconstructed from $U_e = U U_\nu$, with a precision mostly limited by unknown relative phases entering the product. In turn, when the stability principle is applied to the charged lepton sector, $U_e$ largely determines both the charged lepton mass matrix and the right-handed mixing $U_{e^c}$. Therefore, using no theoretical assumption but the stability of the small physical parameters, we are lead in the case of semi-degenerate neutrinos to a well-defined structure for the lepton mass matrices. Interesting features of such a structure are i) the atmospheric angle $\theta_{23}$ and the deviation of the solar angle $\theta_{12}$ from $\pi/4$ are provided by the charged lepton mass matrix in a natural (stable) way; ii) the deviation of $\theta_{12}$ from $\pi/4$ and the size of $\theta_{13}$ turn out to be essentially independent. 

The information obtained on $U_{e^c}$ is particularly useful in the context of SU(5) unification, where $U_{e^c}$ is related to the down quark mixing by Clebsch factors, here constrained non-trivially by stability. We provided a simple example of choice of those factors compatible with stability and leading in first approximation to the relations $m_\tau \approx m_b$, $m_\mu \approx 3 m_s$, $m_e \approx m_d/3$ at the unification scale, in reasonable agreement with data. 

A naturalness argument can also be applied to the neutrino Majorana phases, leading to a mild preference for the Majorana phase $\alpha$ to be close to $\pi/2$. In turn, this has interesting consequences for the mass parameter of neutrinoless double-$\beta$ decay $m_{0\nu\beta\beta}$. In general, the semi-degenerate regime is associated to a large overall neutrino mass scale, which is known to correspond to a experimentally favourable range of $m_{0\nu\beta\beta}$. The above (mild) prejudice on the Majorana phases forces  $m_{0\nu\beta\beta}$ towards the lower edge of the allowed band, corresponding to $m_{0\nu\beta\beta} \approx m \cos2\theta_{12}$. 

A few final remarks. The above considerations would greatly benefit from a better experimental determination of $|U_{32}/U_{31}|$, or equivalently of the Dirac phase  $\delta$, in the standard parameterisation of the PMNS matrix. 
Moreover, while we here considered the case in which the mass matrix entries correspond to independent fundamental parameters, the results can be easily generalised to the case in which they are not independent. This would be the case for example if a non-abelian symmetry correlated different matrix entries.  
Finally, as the neutrino and charged lepton mass matrices are a direct emanation of the physics from which lepton flavour originates, the approach we illustrated may provide pieces of the lepton flavour puzzle, possibly relevant for a bottom-up investigation of the origin of flavour. 

\section*{Acknowledgments}

We thank Serguey Petcov for useful comments and references. The work of A.R.\ was supported by the ERC Advanced Grant no.\ 267985 ``DaMESyFla''. V.D.\ acknowledges the financial support of the UnivEarthS Labex program at Sorbonne Paris Cit\'e (ANR-10-LABX-0023 and ANR-11-IDEX-0005-02) and the Paris Centre for Cosmological Physics, as well as of the European Union FP7-ITN INVISIBLES (Marie Curie Action PITAN-GA-2011-289442-INVISIBLES). V.D.\ moreover thanks SISSA, where large parts of this work were completed, for support and hospitality.

\appendix

\section{The stability condition for $\dm{12} = 0$}
\label{sec:polynomia}

In this Appendix we show how the stability condition leads to simple algebraic conditions on the neutrino mass matrix and in turn to the textures in \eq{stableforms}. 

Let us first consider the $2\times 2$ Majorana case as an illustration. Let $M$ be a $2\times 2$ symmetric complex mass matrix.\footnote{For easier readability, we will suppress the index $\nu$ on the neutrino mass matrix and its eigenvalues in the appendices, $M \equiv M_\nu$ and $m_i \equiv m_i^\nu $.} The physical masses can be obtained as the eigenvalues of $M^\dagger M$, i.e.\ as the solution of a simple quadratic equation:
\begin{equation}
\label{eq:2x2eigen}
m^2_{1,2} = m^2 \pm \frac{\sqrt{\Pi}}{2},
\end{equation}
where 
\begin{equation}
\begin{gathered}
\label{eq:2x2}
2 m^2 = m^2_1+m^2_2 = |M_{11}|^2 + 2 |M_{12}|^2 + |M_{22}|^2 \\
\Pi = (m^2_1-m^2_2)^2 = (|M_{11}|^2 - |M_{22}|^2)^2 + 4 |M_{11}M_{12}^* + M_{12}M^*_{22}|^2 .\end{gathered}
\end{equation}
This shows that the discriminant $\Pi = (m^2_1-m^2_2)^2$ can be expressed as a simple polynomial in $M_{ij}$ and $M_{ij}^*$. Let us now recover the stability condition in the $\dm{12}\to 0$, or $\Pi \to 0$ limit, as in \eq{stabilitylimit}. Stability with respect to variations of the $M_{11}$ element requires  
\begin{equation}
\label{eq:2x2stab11}
M_{11}((|M_{11}+\Delta M_{11}|^2 - |M_{22}|^2)^2 + 4 |M_{11}M_{12}^* + \Delta M_{11}M_{12}^* + M_{12}M^*_{22}|^2) = 0
\end{equation}
for $\Delta M_{11}$ in a neighbourhood of zero. As the expression on the left-hand side above is a polynomial in $\Delta M_{11}$ and $\Delta M^*_{11}$, this requires the coefficient of each $(\Delta M_{11})^n (\Delta M^*_{11})^m$ term (in turn polynomials in $M_{ij}$, $M^*_{ij}$) to vanish. The coefficient of the highest term ($n=m=2$) is $M_{11}$, hence $M_{11} = 0$. Analogously, the highest term in the $M_{22}$ stability condition forces $M_{22} = 0$. The vanishing of $M_{11}$ and $M_{22}$ is then enough to ensure stability, as $M_{ij} \, \Pi(M_{ij} + \Delta M_{ij})$ then vanishes identically for all $ij = 11, 22, 12$. The only texture in which a small $\dm{12}$ is stable is therefore, in the $\dm{12}\to 0$ limit, 
\begin{equation}
\label{eq:2x2stable}
M = m
\begin{pmatrix}
0 & 1 \\
1 & 0
\end{pmatrix} .
\end{equation}
The fact that the previous texture leads, when perturbed, to a small but stable $\dm{12}$ is well known. A precise definition of what ``stable'' means was missing however. Here we have provided such a definition and proven that the above texture is the only stable one. Note that the result is not completely trivial. Had we used the weaker form of the stability condition in which \eq{delta2} is required to hold only for infinitesimal variations of the matrix entries, we would have obtained a different, unsatisfactory result, as such a weaker form is not enough to control the stability. In fact, it is easy to see that any $2\times 2$ matrix with $M_{11}M^*_{12} + M_{12}M^*_{22}$ (if $M_{12}\neq 0$) or $|M_{11}| = |M_{22}|$ (if $M_{12} = 0$) satisfies the weaker condition. For example
\begin{equation}
\label{eq:example} 
M = m
\begin{pmatrix}
1 & 0 \\ 0 & 1
\end{pmatrix} 
\end{equation}
does. On the other hand, as we will see in \App{finite}, $\dm{12}$ is unstable in this case, and the infinitesimal variation misses the instability because the latter develops when the relative variation is small, but larger than $(\dm{12}/(2m^2))^2$. 

\medskip

Let us now get to the $3\times 3$ case and again assume for definiteness that $M$ is symmetric (Majorana). Let us first show that the quantity $\Pi$ in \eq{delta1} is indeed a polynomial in the matrix entries and their conjugated and show how such a polynomial can be calculated. 

The singular values $m_i\geq 0$, conventionally ordered, can be obtained from the eigenvalues $m^2_i$ of $M^\dagger M$. In turn, the latter eigenvalues solve the secular equation $\det(m^2\mathbf{1}_3 - M^\dagger M) = 0$ for $m^2$. The latter is a polynomial equation in $m^2$, as
\begin{equation}
\label{eq:secular}
\det(m^2\mathbf{1}_3 - M^\dagger M) = m^6 - \Pi_1\, m^4 + \Pi_2 \, m^2  - \Pi_3 =
(m^2-m^2_1)(m^2-m^2_2)(m^2-m^2_3)
\end{equation}
where the coefficients $\Pi_{1,2,3}$ are polynomials in $M_{ij}$, $M^*_{ij}$ (whose form can be obtained from \eq{secular}) and in the eigenvalues $m^2_i$
\begin{equation}
\label{eq:Pi123}
\Pi_1 = m^2_1 + m^2_2 + m^2_3,
\quad
\Pi_2 = m^2_1 m^2_2 + m^2_2 m^2_3 + m^2_1 m^2_3,
\quad
\Pi_3 = m^2_1 m^2_2 m^2_3 .
\end{equation}
The expressions for the solutions of the cubic equation (the eigenvalues $m^2_i$) in terms of its coefficients $\Pi_{1,2,3}$ are well known and involve the discriminant
\begin{equation}
\label{eq:Picoeff}
\Pi \equiv 18\, \Pi_1\Pi_2\Pi_3 - 4\, \Pi^3_1\Pi_3 + \Pi_1^2\Pi^2_2 - 4\, \Pi^2_3 - 27\, \Pi_3^2,
\end{equation}
which, when expressed in terms of the eigenvalues $m^2_i$ through \eqs{Pi123}, becomes
\begin{equation}
\label{eq:Pieigen}
\Pi= ((m^2_1-m^2_2)(m^2_2 - m^2_3) (m^2_3 - m^2_1))^2. 
\end{equation}
We therefore see that the combination of eigenvalues in the above equation can be written, through \eq{Picoeff}, as a polynomial in $M_{ij}$, $M^*_{ij}$. The explicit expression is cumbersome and will not be reproduced here. The quantity $\Pi$ can also be obtained (up to a constant) as the lowest order symmetric function of the eigenvalues $m^2_i$ that vanishes if any two eigenvalues coincide.

\Eq{stabilitylimit} gives 
\begin{equation}
0 = |M_{ij}| \, \Pi(M_{ij}+\Delta M_{ij}) = \sum_{nm} |M_{ij}| c_{ij}^{nm}(M_{ij},M^*_{ij}) (\Delta M_{ij})^n (\Delta M_{ij}^*)^m,
\end{equation}
for all $\Delta M_{ij}$ in a neighbourhood of zero. Therefore, $|M_{ij}| c_{ij}^{nm}(M_{ij},M^*_{ij}) = 0$ for all $n,m$ and for each $i,j$. Starting with varying the the off-diagonal elements ($ij = 12,13,23$), an explicit calculation of the leading order coefficients ($n = m = 5$) yields 
\begin{equation}
\label{eq:bce}
\begin{aligned}
|M_{ij}| \, c_{ij}^{55} &= 4 |M_{ij}| (|M_{ii}|^2 + |M_{jj}|^2) && \rightarrow  &&M_{ij} = 0 \text{  or  } M_{ii} = M_{jj} = 0  \quad \text{for} \quad ij = 12,13,23 .
\end{aligned}
\end{equation}
This allows for 4 types of textures,
\begin{equation}
 \begin{pmatrix}
  0 & M_{12} & 0 \\
  M_{12} & 0 & 0 \\
  0 & 0 & M_{33}
 \end{pmatrix} \!, \,
 \begin{pmatrix}
  0 & M_{12} & M_{13} \\
  M_{12} & 0 & 0 \\
  M_{13} & 0 & 0
 \end{pmatrix} \!, \,
  \begin{pmatrix}
  M_{11} & 0 & 0 \\
  0 & M_{22} & 0 \\
  0 & 0 & M_{33}
 \end{pmatrix} \!, \,
 \begin{pmatrix}
  0 & M_{12} & M_{13} \\
  M_{12} & 0 & M_{23} \\
  M_{13} & M_{23} & 0
 \end{pmatrix} 
\end{equation}
and the ones obtained from permutations of rows and columns. Calculating the other coefficients $c^{nm}$ (still for the variation with respect to the off-diagonal elements) and requiring them to be zero eliminates the last texture. 

Turning to the variation of the diagonal elements ($ij = 11,22,33$) and considering again the leading order coefficients, we find
\begin{equation}
\label{eq:diagonal}
\begin{aligned}
 |M_{ii}| \, c_{ii}^{44} &= |M_{ii}| (|M_{jj}|^2 - |M_{kk}|^2)  &&\rightarrow  &&M_{ii} = 0 \text{  or  } |M_{jj}| = |M_{kk}| \,, \\
\end{aligned}
\end{equation}
{with $ijk$ cyclic permuations of 123}.
The remaining textures thus are
\begin{equation}
 \begin{pmatrix}
  0 & M_{12} & 0 \\
  M_{12} & 0 & 0 \\
  0 & 0 & M_{33}
 \end{pmatrix} \!, \,
 \begin{pmatrix}
  0 & M_{12} & M_{13} \\
  M_{12} & 0 & 0 \\
  M_{13} & 0 & 0
 \end{pmatrix} \!, \,
\begin{pmatrix}
  0 & 0 & 0 \\
  0 & 0 & 0 \\
  0 & 0 & M_{33}
 \end{pmatrix} \!, \,
 \begin{pmatrix}
  M_{33} e^{i\beta} & 0 & 0 \\
  0 & M_{33} e^{i \alpha} & 0 \\
  0 & 0 & M_{33}
 \end{pmatrix} .
\label{eq:abcdef}
\end{equation}
The last texture has $\dm{12} = 0$ but also $\dm{23} = 0$ and should therefore be discarded as, for $\dm{23} = 0$, $\dm{12} = 0$ is not equivalent to $\Pi = 0$. Both $\dm{12}$ and $\dm{23}$ are unstable in this texture. The third texture can be obtained from the first one setting $M_{12} = 0$. In order to keep a non zero $\dm{23}$, one parameters in each of the first two textures must be non-zero, while one is allowed to vanish. This leads to the results in \eq{stableforms} and in \Tab{nutextures}.

\section{Stability constraints for finite $\dm{12}$ in texture A}
\label{sec:corrections}

In the realistic case in which $\dm{12}$ is small but not zero, and the neutrino mass spectrum is semi-degenerate, the stability requirement forces the neutrino mass matrix $M$ to be close to the first case in \eq{abcdef}, 
\begin{equation}
 M =  \begin{pmatrix}
  M_{11} &  m & M_{13} \\
   m & M_{22} & M_{23} \\
  M_{13} & M_{23} & m_3
 \end{pmatrix} .
\label{eq:texA}
\end{equation}
where we have assumed a phase convention for the lepton fields in which the dominant 12 and 33 entries $m$ and $m_3$ to be real and positive. The remaining entries $M_{11}$, $M_{22}$, $M_{13}$, $M_{23}$ represent small perturbations. 

The eigenvalues of $M^\dagger M$ can be obtained from a perturbative expansion in the small entries: 
\begin{equation}
\begin{aligned}
m^2_1 & =  m^2 - m\, |M_{11} + M_{22}^*| + \ldots \\
m^2_2 & =  m^2 + m\, |M_{11} + M_{22}^*| + \ldots \\
m^2_3 & = m_3^2 + \ldots .
\label{eq:eigs}
\end{aligned}
\end{equation}
At leading order, this leads for $m_3 > m$ ($m_3 < m$) to normal (inverted) hierarchy, with $\dm{23} \approx m_3^2 - m^2$ and 
\globallabel{eq:dm12pert1}
\begin{align}
\dm{12} &= 2  m\, |M_{11} + M_{22}^*| + \delta (\Delta m^2_{12}), \mytag \label{eq:dm12pert1a} \\
\delta (\Delta m^2_{12}) &= e^{i\gamma}\, \frac{m m_3 (M_{13}^{*2} + M_{23}^{2}) + 2m^2 M_{23} M^*_{13}}{m^2 - m_3^2} + \text{h.c.} + \text{higher orders}  \label{eq:dm12pert1b}, \mytag
\end{align}
where $e^{i\gamma}$ is the phase of $M_{11} + M_{22}^*$. 

We can now impose the stability constraint, \eq{stability}, to $\dm{12}$. Let us begin from the variation with respect to the 11 entry, $M_{11} \to M_{11} + \delta M e^{i\theta}$ (where $\delta M > 0$ and the phase of the variation is factored out). For that, it is enough to use the first term in eq.~(\ref{eq:dm12pert1}a): 
\begin{multline}
\label{eq:stability11}
1 \gtrsim 
\max_\theta \left|
\frac{\Delta (\dm{12})}{\delta M} 
\frac{M_{11}}{\dm{12}} 
\right |
\approx \\
2\frac{|M_{11}|m}{\dm{12}} 
\max_{\theta}
\left|
\frac{|M_{11}+M^*_{22} + \delta M e^{i\theta}| - |M_{11}+M^*_{22}|}{\delta M}
\right|
= 2\frac{|M_{11}|m}{\dm{12}},
\end{multline}
where we have used the fact that the stability inequality must hold for any value of the phase $\theta$. From \eq{stability11}, and the analogous condition for 22 variations, we conclude that 
\begin{equation}
\label{eq:deltaa}
|M_{11}| , |M_{22}| \lesssim 
\frac{\dm{12}}{2m} = \epsilon^2 m. 
\end{equation}
The previous condition ensures that $\dm{12}$ is stable in the leading order approximation. The first contribution to $\dm{12}$ in \eq{dm12pert1}, on the other hand, does not constrain the 13 and 23 entries. Let us then take into account the next to leading order correction in eq.~(\ref{eq:dm12pert1}b) and recover the constraint on $M_{13}$, $M_{23}$. Let us consider first a variation of the 23 element, $M_{23} \to M_{23} + \delta M e^{i\theta}$, which gives 
\begin{multline}
\label{eq:stability23}
1 \gtrsim 
\max_\theta \left|
\frac{\Delta (\dm{12})}{\delta M} 
\frac{M_{23}}{\dm{12}} 
\right |
\approx \\
\frac{|M_{23}|m}{\dm{12}(m^2-m^2_3)}
\max_{\theta}
\left|
2(m_3 M_{23}+ m M^*_{13}) e^{i(\theta+\phi)} + \delta M m_3 e^{i(2\theta+\phi)} + \text{h.c.}
\right| \geq \\
\frac{|M_{23}|m}{\dm{12}(m^2-m^2_3)}
\max \left[
4|m_3 M_{23}+ m M^*_{13}| , 2\,\delta M m_3
\right].
\end{multline}
We therefore have
\begin{equation}
\label{eq:23conditions}
\left\{
\begin{aligned}
& 2\frac{|M_{23}|m}{\dm{12}}\, \frac{\delta M m_3}{|m^2-m^2_3|} \lesssim 1, \\
& 2\frac{|M_{23}|m}{\dm{12}}\, \frac{2|m_3 M_{23}+ m M^*_{13}|}{|m^2-m^2_3|} \lesssim 1,
\end{aligned}
\right.
\end{equation}
for $\delta M \ll M$. Let us now show that the first equation implies 
\begin{equation}
\label{eq:temp1}
R\equiv \fracwithdelims{(}{)}{2 m m_3 |M_{23}|^2}{\dm{12}|m^2-m^2_3|}^{1/2} \lesssim 1. 
\end{equation}
If this was not the case, i.e.\ if $R\gg 1$, we could consider a variation of $M_{23}$ of size $\delta M = |M_{23}| /R \ll |M_{23}|$, for which we would have
\begin{equation}
\label{eq:temp2}
2\frac{|M_{23}|m}{\dm{12}}\, \frac{\delta M m_3}{|m^2-m^2_3|} = \frac{\delta M}{|M_{23}|} R^2 = R \gg 1,
\end{equation}
in contradiction with the first condition in \eq{23conditions}. The stability constraint on $M_{23}$ (and, analogously, the one on $M_{13}$) follows from \eq{temp1}:
\begin{equation}
\label{eq:23constr}
|M_{13}|, |M_{23}| \lesssim 
\left(
\dm{12} \frac{|m^2-m^2_3|}{2m m_3} 
\right)^{1/2}
= \epsilon k m
\end{equation}
Using this result, one also gets a bound on the product $M_{13}M_{23}$ from the second condition in \eq{23conditions}:
\begin{equation}
\label{eq:2313constr}
|M_{13}M_{23}|^{1/2} \lesssim 
\epsilon k \sqrt{mm_3} \sim \epsilon k m .
\end{equation}
In summary, the neutrino mass matrix is constrained by stability to be in the form in \eq{SDtexture}. We have explicitly checked that that is also a sufficient condition for stability. Note that given the bounds eq.~\eqref{eq:deltaa} and eq.~\eqref{eq:23constr}, the two contributions in eq.~\eqref{eq:dm12pert1a} and \eqref{eq:dm12pert1b} turn out to be of the same order in $\epsilon$. Using this information and rederiving the above constraints order by order in $\epsilon$ confirms the bounds on the elements of $M_\nu$ derived in this section, proving the self-consistency of this analysis.

We close this appendix discussing how the above results change for texture B, i.e.\ when $m_3$ is set to zero. The bounds on the 11 and 22 elements of $\Delta M$ do not depend on $m_3$ and therefore do not change. The 33 element of $\Delta M$, on the other hand, is now allowed to be sizeable, as we know from the stability of texture A. On the other hand, a sizeable 33 element brings us back to texture A. We can therefore say that the 33 element of $\Delta M$ is small by the very definition of texture B. Similar considerations hold for the 13 and 23 elements. While their product is still bounded (by $\epsilon^2 m^2/2$, see the second condition in \eq{23conditions}), the individual elements $\Delta M_{13}$ and $\Delta M_{23}$ are now allowed to be sizeable, provided that the other one is correspondingly suppressed. However, a sizeable $\Delta M_{13}$ or $\Delta M_{23}$ brings us towards texture C. We can therefore again say that  the 13 and 23 elements of $\Delta M$ are small by the very definition of texture B. In the end, we get for texture B results similar to those found for texture A, i.e.\ $\lesssim \epsilon^2$ deviations from $\pi/4$ for the 12 rotation and $\lesssim \epsilon$ contributions in $U'$ in \eq{SDtexture}.

\section{Finite differences against infinitesimal variations}
\label{sec:finite}

It is instructive to consider again the $2\times 2$ case, which nicely shows why the infinitesimal form of the stability condition is not enough to exclude the texture in \eq{example}. Let us add a small, off-diagonal element to that texture in order to generate a small $\dm{12}$:
\begin{equation}
\label{eq:examplefinite}
M = m
\begin{pmatrix}
1 & \epsilon \\
\epsilon & 1
\end{pmatrix},
\end{equation}
with $0<\epsilon\ll 1$ (not to be confused with the $\epsilon$ in \eq{semideg2}). We then have $  2\epsilon \approx \dm{12}/(2m^2)\ll 1$. Let us now study the behaviour of $\Pi = (\dm{12})^2$\footnote{As in other cases, we consider $(\dm{12})^2$ instead of $\dm{12}$ simply because $\Pi$ has a polynomial expression in the matrix entries that turns useful when computing finite variations.} with respect to (real) variations of the matrix entries. When using infinitesimal variations we get
\begin{equation}
\label{eq:dPi}
\left|
\frac{\partial\Pi}{\partial M_{11}} \frac{M_{11}}{\Pi}
\right| = 
\left|
\frac{\partial\Pi}{\partial M_{22}} \frac{M_{22}}{\Pi}
\right| = 1, \qquad
\left|
\frac{\partial\Pi}{\partial M_{12}} \frac{M_{12}}{\Pi}
\right| = 2 .
\end{equation}
The texture appears to be stable. But this is not the case. Let us consider now a variation of the entries by a finite amount $\delta$ ($1\to 1+\delta$, or $\epsilon \to \epsilon + \delta$). We now have
\begin{equation}
\label{eq:deltaPi}
\left|
\frac{\Delta\Pi}{\Delta M_{11}} \frac{M_{11}}{\Pi}
\right| = 
\left|
\frac{\Delta\Pi}{\Delta M_{22}} \frac{M_{22}}{\Pi}
\right| = 1 + \delta \left( \frac{1}{4} + \frac{1}{4\epsilon^2} \right) + \frac{\delta^2}{4\epsilon^2} + \frac{\delta^3}{16\epsilon^2},
\quad
\left|
\frac{\Delta\Pi}{\Delta M_{12}} \frac{M_{12}}{\Pi}
\right| = 2 + \frac{\delta}{\epsilon}.
\end{equation}
The infinitesimal limit is recovered when $\delta \ll 4\epsilon^2 \ll  (\dm{12}/(2m^2))^2$. On the other hand, when $4\epsilon^2 \ll \delta \ll 1$, the instability emerges, 
\begin{equation}
\label{eq:deltaPiagain}
\left|
\frac{\Delta\Pi}{\Delta M_{11}} \frac{M_{11}}{\Pi}
\right| \approx
\frac{\delta}{4\epsilon^2} \gg 1 .
\end{equation}
A finite variation larger than the (square of the) small scale of the problem, $(\dm{12}/(2m^2))^2$, is necessary in order to see the instability. This is similar to what was found in~\cite{Marzocca:2014tga} for charged leptons. 

\section{Stability of the charged lepton mass matrix}
\label{sec:stabproof}

In this Appendix we prove the statements in eqs.~(\ref{eq:ME1},\ref{eq:Ucstability}). We will make use of Proposition 2 in~\cite{Marzocca:2014tga}. According to which, the stability of $M_E$ is equivalent to 
\begin{equation}
\begin{gathered}
|M^E_{ih} M^E_{jk}| \lesssim m_\mu m_\tau \quad \text{for all $i\neq j$, $h\neq k$} \\
|M^E_{1i} M^E_{2j} M^E_{3k}| \lesssim m_e m_\mu m_\tau \quad \text{for all $ijk$ permutations of $123$} .
\end{gathered}
\label{eq:proposition}
\end{equation}
We order the singlet leptons $e^c_i$ in such a way that $|U^{e^c}_{ii}| \sim 1$ and denote 
\begin{equation}
t_e = \tan\theta^e_{12}, \quad t' = \tan\theta'_{12}, \quad t_\text{max} = \max(t_e,t'), \quad t_\text{min} = \min(t_e,t'). 
\label{eq:notation}
\end{equation}

According to \eq{thirdrow}, the third row of $M_E$ is in the form $(|M^E_{3i}|) \sim (t_e m_\tau, m_\tau, m_\tau)$. Here we are neglecting the $\lesssim\epsilon$ contributions from $U'$ in \eq{SDtexture}, a point we will return to in the second part of this Appendix. Stability then requires $|M^E_{2i}| \lesssim m_\mu$ (same for for the first row $M^E_{1i}$). Then
\begin{equation}
m_\mu \gtrsim |M^E_{23}| = |U^{e^c}_{32} m_\tau U^e_{33} + \ord{\leq m_\mu}|, 
\label{eq:app1}
\end{equation}
together with $|U^e_{33}|\sim 1$, implies $|U^{e^c}_{32}| \lesssim m_\mu/m_\tau$. Analogously, $|U^{e^c}_{31}| \lesssim m_\mu/m_\tau$. Using the latter result and the explicit form of $U_e$ in the expression $M^E_{21} = U^{e^c}_{k2} m_k U^e_{k1}$, one finds that $|M^E_{21}| \lesssim t_\text{max} m_\mu$. Moreover, at least one out of $|M^E_{22}|$ and $|M^E_{23}|$ must be of order $m_\mu$. This follows from 
\begin{equation}
\begin{aligned}
U^{e^c}_{22} m_\mu & = (U^e_{33} M^E_{22} - U^e_{32} M^E_{23})/D +\ord{m_e} ,
\end{aligned}
\label{eq:app2}
\end{equation}
and $|U^{e^c}_{22}| \sim 1$, where $D$ is the determinant of the 23 block of the matrix $U_e$, $D=e^{-i\phi_e} c^e_{12}c'_{12} - \hat c_{23} s^e_{12} s'_{12} \approx e^{-i\phi_e} c^e_{12} c'_{12} \sim 1$. In the following we will assume for definiteness that $|M^E_{23}|\sim m_\mu$. The results for the case in which $|M^E_{22}| \sim m_\mu$ can be obtained by exchanging the last two columns of $M_E$. This is the origin of the permutation matrix $P_{23}$ in \eq{ME1}.  All in all, the second line of $M_E$ must then be in the form $(|M^E_{2i}|) = (\lesssim t_ \text{max} m_\mu, \lesssim m_\mu, \sim m_\mu)$. 

Let us now consider the first row of $M_E$. \Eq{proposition} requires $|M^E_{11}| \lesssim m_e$, $|M^E_{12}| \lesssim m_e/t_e$, and inverting $M_E = U_{e^c}^T M^\text{diag}_E U_e$ we obtain
\begin{equation}
|U^{e*}_{11}M^E_{11} + U^{e*}_{12}M^E_{12} + U^{e*}_{13}M^E_{13}|/m_e = |U^{e^c}_{11}| \leq 1, 
\label{eq:app3}
\end{equation}
which forces $|M^E_{13}| \lesssim m_e/t_\text{min}$. Therefore we have, for the first row of $M_E$, $(|M^E_{1i}|) = (\lesssim m_e, \lesssim m_e /t_e, \lesssim m_e/t_ \text{min})$. We can still improve on the above approximate bound. Using
\begin{equation}
m_e \gtrsim |M^E_{11} c^e_{12} + M^E_{12} s^e_{12} | = |m_e c'_{12} U^{e^c}_{11} -m_\mu s'_{12} U^{e^c}_{21}|
\label{eq:app4}
\end{equation}
we get $|U^{e^c}_{21}| \lesssim m_e/(m_\mu t')$, and using
\begin{multline}
1 \gtrsim 
\frac{|M^E_{13} M^E_{21} M^E_{32} - M^E_{13} M^E_{22} M^E_{31}|}{m_e m_\mu m_\tau} \\
\approx \left|U^{e^c}_{22} U^{e^c}_{33} s'_{12} \hat s_{23} 
\left[
U^{e^c}_{31} \hat c_{23} \frac{m_\tau}{m_e}+e^{-i\phi_e} \hat s_{23}
\left(
U^{e^c}_{21} c'_{12} \frac{m_\mu}{m_e} + U^{e^c}_{11} s'_{12}
\right)
\right]\right|
\label{eq:app5}
\end{multline}
we get $|U^{e^c}_{31}| \lesssim m_e/(m_\tau t')$. Using the above bounds in the expressions $M^E_{1j} = U^{e^c}_{k1} m_k U^e_{kj}$ for the matrix elements of the first row of $M_E$ we obtain the bounds $|M^E_{12}| \lesssim m_e/t' $, $|M^E_{13}| \lesssim m_e/t'$ that, together with the previous ones, give $(|M^E_{1i}|) = (\lesssim m_e, \lesssim m_e /t_\text{max}, \lesssim m_e/t')$. 

All in all we get the following stability bounds on the charged lepton mass matrix
\begin{equation}
|M_E| =
\begin{pmatrix}
\lesssim m_e & \lesssim m_e/t_\text{max} &  \lesssim m_e/t' \\
\lesssim m_\mu t_\text{max} & \lesssim m_\mu & \sim m_\mu \\
\sim m_\tau t_e & \sim m_\tau & m_\tau
\end{pmatrix}, \quad\text{besides}\quad
\label{eq:appME}
\begin{aligned}
|M^E_{12} M^E_{21} | &\lesssim m_e m_\mu \\
|M^E_{13} M^E_{21} | &\lesssim m_e m_\mu \\
|M^E_{13} M^E_{22} | &\lesssim m_e m_\mu /t_e
\end{aligned} .
\end{equation}
We can now show that the above bounds are equivalent to the existence of a $t$ in the range $t' \leq t \leq \max(t',t_e)$ satisfying the bounds in \eq{ME1}. Clearly, if $M_E$ satisfies the bounds in \eq{ME1}, with $t$ in the above range, then it also satisfies the bounds in \eq{appME}. In order to show the the vice versa also holds, we observe that \eq{appME} implies the following 9 bounds
\begin{equation}
t', \frac{|M^E_{21}|}{m_\mu}, \frac{|M^E_{22}|t_e}{m_\mu} \lesssim 
\frac{m_e}{|M^E_{21}|}, \frac{m_e}{|M^E_{13}|}, t_\text{max} .
\label{eq:appbounds}
\end{equation}
It then suffices to choose $t$ such that 
\begin{equation}
t' \leq \max \left(t', \frac{|M^E_{21}|}{m_\mu}, \frac{|M^E_{22}|t_e}{m_\mu} \right)
\lesssim t \lesssim
\min \left(
\frac{m_e}{|M^E_{21}|}, \frac{m_e}{|M^E_{13}|}, t_\text{max} 
\right) \leq t_\text{max}  
\label{eq:appbounds2}
\end{equation}
(and to make sure that $t' \leq t \leq t_\text{max}$, with no wiggles) in order to satisfy the bounds in \eq{ME1}. 

Finally, \eq{Ucstability} follows from using the bounds in \eq{ME1} in the expression $U^{e^c}_{ki} = (U^e_{kj})^* M^E_{ij}/m_k$, obtained inverting $M_E = U_{e^c}^T M^\text{diag}_E U_e$. 

Now, let us return to the $\lesssim\epsilon$ contributions from the 13- and 23-rotations by angles $\theta^\nu_{13}$ and $\theta^\nu_{23}$ in the neutrino sector, leading to $\lesssim\epsilon$ contributions in all entries of $U_e$. This can affect the above results wherever we use the explicit form of $U_e$. Most importantly, 
\begin{align}
 U_{31}^e &  = \sin(\hat\theta_{23}) \sin(\theta_{12}^e) - \cos(\hat\theta_{23}) \, \epsilon_{13} \label{eq:U31e} \,,\\
 U_{13}^e & = \hat s_{23} s_{12}' + c_{12}' e^{i  \phi_e}( \hat c_{12}  \epsilon_- - i \hat s_{12}  \epsilon_+) -  \hat c_{23} s_{12}'( i \hat c_{12} \epsilon_+  + \hat s_{12} \epsilon_-) \,, \label{eq:U13e}
\end{align}
with $\epsilon_{13} = \sin(\theta_{13}^\nu) e^{- i \varphi_{13}^\nu}$, $\epsilon_+ = e^{i \beta} (\sin(\theta_{13}^\nu) e^{i \varphi_{13}^\nu} + \sin(\theta_{23}^\nu) e^{i \varphi_{23}^\nu})/\sqrt{2}$, $\epsilon_- = e^{i \alpha} (\sin(\theta_{13}^\nu) e^{i \varphi_{13}^\nu} - \sin(\theta_{23}^\nu) e^{i \varphi_{23}^\nu})/\sqrt{2}$ and $|\epsilon_{13}|, |\epsilon_{\pm}|\lesssim \epsilon$. Hence, with $\epsilon \leq 0.09$ for $m_1 > 0.05$~eV, cancellations between the ${\cal O}(\theta_{12}^\nu)$ terms and the  ${\cal O}(t_e, t')$ terms are possible. This implies that strictly speaking there is no lower bound on $U_{31}^e$, $U_{13}^e$, unlike the case in which the $\epsilon_{13,\pm}$ corrections can be neglected. On the other hand, a vanishing value for $|U^e_{31,13}|$ is not the generic case, but occurs only for specific parameter choices. In particular, the $\epsilon_{13,\pm}$ corrections have to be sufficiently close to the upper bounds of eq.~\eqref{eq:SDtexture}. Still, if this is the case and the cancellation takes place, the bound on the first row of $M_E$ is significantly weakened, 
\begin{equation}
 |M_E| = \begin{pmatrix}
  \lesssim m_e & \lesssim m_\mu & \lesssim m_\mu \\
  \lesssim m_\mu  t'_\text{max}& \lesssim m_\mu & \sim m_\mu \\
  \ll m_\tau & \sim m_\tau & \sim m_\tau
 \end{pmatrix} \,.
 \label{eq:MEepsilon}
\end{equation}
with $t'_\text{max} = \text{max}(t_\text{max}, \tan(\theta^\nu_{13}))$. For $U_{e^c}$ this implies
\begin{equation}
 |U_{e^c}| = \begin{pmatrix}
              \sim 1 & < 1 & \lesssim m_\mu/m_\tau  \\
              < 1 & \sim 1 & \lesssim m_\mu/m_\tau \\
           \lesssim m_\mu/m_\tau  &  \lesssim m_\mu/m_\tau &  \approx 1
               \end{pmatrix} \,.
               \label{eq:Uecepsilon}
\end{equation}
In the main part of this paper, we focus on the situation in which this cancellation does not occur - relevant for the vast part of the parameter space. This is however a special situation to be kept in mind as it allows to evade some of the bounds imposed on the structure of the charged lepton mass matrix and mixing.

\bibliographystyle{JHEP}
\bibliography{ref}

% \bibliographystyle{h-physrev4}
% \bibliography{abbrev,biblio,hep,tmp,pro}

\end{document}